\documentclass[12pt]{article}

\usepackage{latexsym}

\usepackage{amsfonts}
\usepackage{amssymb}

\font\tenmsbm=msbm10 scaled 1200
\font\sevenmsbm=msbm9
\newfam\msbmfam
\textfont\msbmfam=\tenmsbm \scriptfont\msbmfam=\sevenmsbm

\makeatletter
\@addtoreset{equation}{section}
\makeatother

\newcommand{\eref}[1]{(\ref{#1})}

\def\be{\begin{equation}}
\def\ee{\end{equation}}
\def\ba{\begin{eqnarray}}
\def\ea{\end{eqnarray}}
\def\bet{\begin{tabular}}
\def\eet{\end{tabular}}
\def\pa{\partial}

\def\nn{\nonumber}
\def\ve{\varepsilon}

\def\a{\alpha}
\def\bt{\beta}
\def\bt{\beta}
\def\G{\Gamma}
\def\D{\Delta}
\def\dl{\delta}
\def\dl{\delta}

\def\s{\sigma}
\def\m{\mu}
\def\n{\nu}
\def\Th{\Theta}
\def\ra{\rightarrow}
\def\vp{\varphi}
\def\ul{\underline}
\def\wt{\widetilde}
\def\wh{\widehat}

\catcode`@=11

\begin{document}

\begin{titlepage}

\begin{flushright}
Preprint DFPD/2010/TH16\\
November 2010\\
\end{flushright}

\vspace{2.5truecm}

\begin{center}

{\Large \bf Ultraviolet singularities in classical brane
theory} \vskip0.3truecm

\vspace{2.0cm}

K. Lechner\footnote{kurt.lechner@pd.infn.it}

\vspace{2cm}

 {
\it Dipartimento di Fisica, Universit\`a degli Studi di Padova, Italy

\smallskip

and

\smallskip

INFN, Sezione di Padova,

Via F. Marzolo, 8, 35131 Padova, Italy
}

\vspace{1.5cm}

\begin{abstract}

We construct for the first time an energy--momentum tensor for the electromagnetic field of a $p$--brane in arbitrary dimensions, entailing finite energy--momentum integrals.
The construction relies on distribution theory and is based on a Lorentz--invariant regularization, followed by the subtraction of divergent and finite counterterms supported on the brane. The resulting energy--momentum tensor turns out to be uniquely determined.  We perform the construction explicitly for a generic flat brane. For a brane in arbitrary motion our approach provides a new paradigm for the derivation of the, otherwise divergent, self--force of the brane. The so derived self--force is automatically finite and guarantees, by construction, energy--momentum conservation.

\vspace{0.5cm}

\end{abstract}

\end{center}
\end{titlepage}

\newpage

\baselineskip 6 mm
%


\section{Introduction}

Like classical Electrodynamics of charged point--particles in four dimensions, the classical theory of charged extended objects, or branes, in arbitrary dimensions is plagued by ultraviolet singularities that make the theory -- as it stands -- inconsistent.
There are two types of singularities showing up -- both caused by the singular behavior of the electromagnetic field in the vicinity of the brane --  that are, however, a priori unrelated to each other: \\
I) {\it The energy--momentum of the electromagnetic field in a volume enclosing (a portion of) the brane is infinite.}
\\
II) {\it The self--force experienced by the brane is infinite.}

The purpose of this paper is to provide a universal approach that, a) eliminates the divergences of the first type, and b) entails automatically a finite self--force, so that divergences of the second type will never show up. The main motivation for this development is that the self--force and the energy--momentum integrals are
crucial ingredients of brane--radiation theory, that is, of a systematic analysis of the radiation emitted from, and the back reaction experienced by, a brane. A part from their conceptual relevance, these phenomena are of interest, for example, in cosmological models based on brane inflation and cosmic superstrings  \cite{S}.

The guiding principle of our approach will be energy--momentum conservation -- a principle that in a relativistic theory requires the construction of a well--defined, and conserved, energy--momentum tensor. In the case of $\dl$--like sources -- like point--particles and branes -- the main problem with this respect is that, due
to the singular behavior of the  electromagnetic field near the sources, the standard energy--momentum tensor of the field {\it is not a distribution}: it is this circumstance that causes, eventually, the divergences of the type I). One of the main achievements of this paper is the development of a general approach for the construction of a well--defined energy--momentum tensor for the field of a brane. To overcome the divergences of the type I) this tensor must thus live necessarily in the space of distributions.
The approach we propose is based on a regularization that preserves Lorentz--invariance in target--space as well as diffeomorphism--invariance on the brane worldvolume, followed by a classical renormalization involving {\it divergent} and {\it finite} counterterms localized on the brane.

Standard derivations  of the self--force, see e.g. \cite{KLS}--\cite{KS}, are usually based on some regularization of the infinite self--force, with the aim of isolating its divergent part. These procedures entail as main drawback --  a part from the unclear fate of the divergent part --  the lack of control over energy--momentum conservation. This last feature is particularly problematic, since the resulting equations of motion can not be deduced from an action \cite{KLS,G2}, and N\"other's theorem can therefore not be applied. In our approach the derivation of the self--force is based, instead, ab initio on energy--momentum conservation -- but realized in the distributional sense -- and it leads directly to a finite self--force, so that divergences of the type II) never arise. In particular, the so obtained energy--momentum tensors and self--forces turn out to be uniquely determined.

The approach we propose applies in principle to a generic brane in arbitrary motion, but in this paper we give a constructive proof that it works for an arbitrary {\it flat brane}, that is, a rigid brane in uniform motion, for which the self--force must vanish. It generalizes a method that has been applied successfully to charged particles \cite{LM} and dyons \cite{L} in arbitrary motion in four dimensions, and we hope to prove its full efficiency for a brane in arbitrary motion elsewhere.

In this paper we limit ourselves to branes coupled minimally to an antisymmetric potential, but our approach applies equally well if they are coupled to scalar fields, or to a (linearized) gravitational field.

The material is organized as follows.
In the next subsection we illustrate the
ultraviolet divergences I), II), and their taming, in the case of a charged particle in $D=4$, summarizing the results of \cite{LM}. In section \ref{pbrane} we present our new approach for a generic brane in $D$ dimensions. Section \ref{stat} is devoted to a description of the dynamics of flat branes, to which we will concentrate mainly in the remaining sections. In sections \ref{p6} and \ref{s4} we illustrate the approach for the relatively simple cases of a particle in uniform motion in $D=6$, and a flat string in $D=4$. These cases are
prototypical and exhibit already the main characteristic features of our approach.
In section \ref{general} we give a constructive proof of its efficiency for an arbitrary flat brane in $D$ dimensions. Section \ref{dyn} is devoted to a preliminary analysis of branes in arbitrary motion, while section \ref{sum} contains a brief summary and lists the open problems. Some technical details are relegated to two appendices.

\subsection{Ultraviolet divergences for a particle in $D=4$}\label{udf}

Both divergences mentioned above originate in general from the singular behavior of the generalized electromagnetic field,
\be\label{fp}
F_{\m_1\cdots\,\m_{p+1}}= (p+1)\,\pa_{[\m_1} A_{\m_2\cdots\,\m_{p+1}]},
\ee
in the vicinity of the brane. This expression refers to a $(p-1)$--brane, for which the generalized potential $A_{\m_1\cdots\,\m_p}$ is an antisymmetric tensor of rank $p$.
For a particle in $D=4$ we have $p=1$,
and the dynamics is governed by the Lorentz and Maxwell equations,
\ba
\label{lor}
M{dU^\m\over d\s}&=&e\, F^{\m\n}(x(\s))U_\n,\\
 \pa_\m F^{\m\n}&=&J^\n=e\int U^\n\dl^4(x-x(\s))\,d\s.\label{max}
\ea
Here $\s$ is the proper time, $x^\m(\s)$ parametrizes the worldline, $U^\m=dx^\m(\s)/d\s$ is the four--velocity, and in the Lorentz equation $F^{\m\n}$ is evaluated on the worldline of the particle. In the vicinity of the particle the field
diverges, schematically, as
\be\label{fmn}
F^{\m\n}(x)\sim 1/r^2,
\ee
with $r=|\vec x-\vec x(\s)|$, and  consequently the ``bare'' self--force, $$
eF^{\m\n}(x(\s))U_\n,
$$
is infinite. Here and in the following we ignore external fields, so that $F^{\m\n}$ is just the Lienard--Wiechert field. As it stands, the Lorentz--equation is therefore meaningless.

A standard strategy to attack this problem consists in regularizing the field $F^{\m\n}$ in some way, to evaluate the regularized field on the worldline, and to send  then the regulator to zero. In doing so the right hand side of \eref{lor} develops a divergent part, that can be absorbed by an (infinite) mass shift $M\rightarrow M+\Delta M$, and a finite part. Keeping only the latter, the ill--defined equation \eref{lor} goes over to the Lorentz--Dirac equation \cite{D},
\be
\label{ld}
M{dU^\m\over d\s}={e^2\over 6\pi}\left({d^2U^\m\over d\s^2}+\left({dU\over d\s}\right)^2U^\m\right).
\ee
In this paper we use the mostly minus signature $(+,-,\cdots,-)$.
The right hand side of this equation identifies the ``self--force'' of the particle.
The Lorentz--Dirac equation represents, on one hand, a cornerstone of classical radiation theory -- that quantifies the self--interaction of a charged particle  -- but, on the other hand, it inherits several unpleasant features: a) it can not be {\it derived} from the fundamental equations \eref{lor}, \eref{max}, and eventually it must be {\it postulated}; b) it is of {\it third} order in time derivatives and hence in conflict with Newton's determinism; c) it is Lorentz--covariant, but it can not be derived from an action, and hence four--momentum conservation is  -- a priori -- not guaranteed. It is in particular feature b) that turns classical Electrodynamics of point--particles into an internally inconsistent theory. Nevertheless, from an experimental point of view equation \eref{ld} describes correctly the emission of four--moment due to radiation, up to the quantum energy scale, i.e. for wavelengths $\lambda\gg \hbar/Mc$ \cite{R}, and therefore an independent criterion to establish its ``validity'' should be pursued.

One of the purposes of the present paper is to provide such a general and feasible criterion, that applies to a generic brane in arbitrary dimensions.
As noted already by Dirac in \cite{D}, the principal clue for the justification of \eref{ld} arises from energy conservation, a principle that in a relativistic theory we realize through a conserved total energy--momentum tensor. It is at this point that the divergence of the type I) comes into the game: since the ``bare'' energy--momentum tensor of the field is
$\Th^{\m\n}= F^{\m\a}F_\a{}^\n+{1\over4}\,\eta^{\m\n}F_{\a\bt}F^{\a\bt}$,
due to \eref{fmn} the four--momentum density of the field diverges near the particle as,
\be\label{tdiv}
\Th^{0\m}\sim 1/r^4,
\ee
and hence the energy $E_V=\int_V \Th^{00}d^3r$, and more generally the four--momentum $P^\m_V=\int_V \Th^{0\m}d^3r$, are {\it infinite} if the particle stays inside the volume $V$. As it stands, the concept of four--momentum conservation is thus meaningless.

From a mathematical point of view this problem originates from the fact that $\Th^{\m\n}$ is not a  distribution; indeed, due to \eref{tdiv} it is not locally integrable \footnote{While $F^{\m\n}$ is a distribution, its products -- like the terms in  $\Th^{\m\n}$ -- are in general not distributions.}. As a consequence also the four--divergence $\pa_\m \Th^{\m\n}$ is ill--defined, because the derivative of a function that is not a distribution is in general not defined. Indeed, the standard ``naive'' calculation,
\be\label{dtmn}
\pa_\m \Th^{\m\n}=F^{\m\n}J_\m =e\int F^{\m\n}(x(\s))\,U_\m\,\dl^4(x-x(\s))\,d\s,
\ee
does not make sense, since the self--field $F^{\m\n}(x(\s))$ is infinite. The question whether the total energy--momentum tensor is a conserved tensor is, therefore, ill--posed.

\ul{\it Renormalized energy--momentum tensor.}
The strategy to justify \eref{ld} through four--momentum conservation requires, therefore, an energy--momentum tensor for the electromagnetic field, say $T^{\m\n}$ instead of $\Th^{\m\n}$, that A) is  a distribution, and B) does not differ ``too much'' from $\Th^{\m\n}$, in the sense that away from the worldline, i.e. in its complement, the  tensors must coincide, $T^{\m\n}= \Th^{\m\n}$. We call $T^{\m\n}$ the ``renormalized energy--momentum tensor'' of the field. For a particle in four dimensions a $T^{\m\n}$ with these properties has been constructed for the first time in a -- pioneering -- paper by P. Rowe in \cite{Ro}, using a somewhat cumbersome and implicit distribution
technique, that relied on the peculiar properties of the Lienard--Wiechert fields in four dimensions. A physically more transparent, and conceptually more simple, construction of $T^{\m\n}$, always for a particle in $D=4$, has been presented in \cite{LM}, where it has also been shown that the approaches of \cite{LM} and \cite{Ro} lead to the same renormalized energy--momentum tensor. Actually, with the requirements A) and B), and demanding the self--force to be ``algebraic'', see paragraph \ref{cc},
the renormalized  energy--momentum tensor turns out to be {\it unique} \cite{Ro}.

\subsubsection{Construction of the renormalized energy--momentum tensor and derivation of the self--force}

The method adopted in \cite{LM} to construct a renormalized energy--momentum tensor consists in isolating and subtracting from $\Th^{\m\n}$ the singularities present along the worldline in a Lorentz--covariant way, {\it without modifying the values of $\Th^{\m\n}$ in the complement of the worldline}. Technically it relies on a procedure very common in quantum field theory: a Lorentz--covariant regularization, followed by the subtraction of singular local terms. With ``local'' we mean here ``supported on the worldline''. A basic asset of this method is that, in conjunction with energy--momentum conservation, it provides a {\it derivation} of the Lorentz--Dirac equation, and hence of the self--force.

The method proceeds along the following steps. Replace the Lienard--Wiechert potential $A^\m$, where the retarded time $\s(x)$ is determined by the standard conditions $(x-x(\s))^2=0$, $x^0-x^0(\s)>0$, by a regularized Lienard--Wiechert potential $A^\m_\ve$, in which the regularized retarded time $\s_\ve(x)$ is determined by,
\be
(x-x(\s))^2=\ve^2, \quad x^0-x^0(\s)>0,
\ee
where $\ve$ is a positive regulator with the dimension of length. This potential can be written in terms of the regularized kernel of the d'Alembertian,
\be\label{ge4}
G_\ve\equiv{1\over 2\pi}H(x^0)\,\dl(x^2-\ve^2),
\ee
as,
\be\label{amue}
A^\m_\ve(x)=G_\ve*J^\m=\left.{e\over 4\pi} {U^\m (\s)
\over (x_\n - x_\n(\s)) U^\n(\s)} \right|_{\displaystyle\,\s=\s_\ve(x)},
\ee
where $H$ denotes Heaviside's step function.
The regularized potential \eref{amue} has been introduced for the first time in \cite{Da}.
It is easily seen that $A^\m_\ve$ is regular on the worldline as long as $\ve>0$.
Introduce then the regularized field $F^{\m\n}_\ve=\pa^\m A^\n_\ve-\pa^\n A^\m_\ve$, and the regularized energy--momentum tensor,
\be\label{treg}
\Th^{\m\n}_\ve=F^{\m\a}_\ve F_{\ve\,\a}{}^\n+{1\over4}\,\eta^{\m\n}F_{\ve\,\a\bt}F_\ve^{\a\bt}.
\ee
Define, eventually, the distribution--valued renormalized energy--momentum tensor,
\ba\label{p4}
T^{\m\n}&=&{\rm Lim}\,_{\ve\ra 0}\left(\Th^{\m\n}_\ve-\wh\Th^{\m\n}_\ve\right),\\
\wh\Th^{\m\n}_\ve&\equiv&
{e^2\over 32\ve}\int\left(U^\m U^\n-{1\over 4}\,\eta^{\m\n}\right)\dl^4(x-x(\s))\,d\s,\label{p4div}
\ea
where, from now on, with the capital ``${\rm Lim}\,_{\ve\ra 0}$'' we mean  ``{\it limit for $\ve\ra 0$ in the sense of distributions}''. In \eref{p4} we subtracted the divergent local ``counterterm'' $\wh\Th^{\m\n}_\ve$ -- proportional to the pole $1/\ve$  -- that cancels from
$\Th^{\m\n}_\ve$
the terms that diverge in the distributional sense as $\ve\ra0$: while the structure of $\wh\Th^{\m\n}_\ve$  is fixed by locality, Lorentz--invariance, and dimensionality, its coefficients are fixed by this cancelation requirement. The crucial point is the following: while for $\ve\ra 0$ the functions $\Th^{\m\n}_\ve(x)$ converge in the complement of the worldline {\it pointwise} to $\Th^{\m\n}(x)$, $\Th^{\m\n}_\ve$ does not converge to the former {\it in the sense of distributions}. This feature will be illustrated in explicit examples in sections \ref{p6} and \ref{s4}. In reference \cite{LM} it has been proven, in particular, that:
\\
a) The distributional limit in \eref{p4} exists, and hence $T^{\m\n}$ is a distribution.
\\
b) $T^{\m\n}$ is a Lorentz--covariant, symmetric and traceless tensor.
\\
c) In the complement of the worldline $T^{\m\n}=\Th^{\m\n}$.
\\
d) The four--divergence of $T^{\m\n}$ equals,
\ba\label{divtem}
 \pa_\m T^{\mu\nu}&=&- \int f^\n\,\delta^4(x-x(\s))\,d\s,\\
f^\n&\equiv& {e^2\over 6\pi}\left({d^2U^\n\over d\s^2} +\left({dU\over d\s}\right)^2
U^\n\right),\label{fvp4}
 \ea
for an arbitrary worldline $x^\m(\s)$.

Once property a) has been ascertained, property b) is obvious. Property c) follows form the fact that the counterterm $\wh\Th^{\m\n}_\ve$ is local, i.e. supported on the worldline. Property a) ensures that the four--divergence $\pa_\m T^{\mu\nu}$ is well--defined, and property c) implies then that it is supported on the worldline, because away from the worldline $T^{\m\n}=\Th^{\m\n}$, and there $\pa_\m\Th^{\m\n}$ is zero, see \eref{dtmn}. This means that $\pa_\m T^{\mu\nu}$ has necessarily the structure \eref{divtem} for some four--vector $f^\n$, that is to become the self--force. An explicit evaluation of $\pa_\m T^{\mu\nu}$ from the definition \eref{p4} gives eventually \eref{fvp4}.

Introducing the energy--momentum tensor of the particle $t^{\m\n}=M\int U^\m U^\n\dl^4(x-x(\s))\,d\s$, from \eref{divtem} one concludes that the divergence of the total energy--momentum tensor $\tau^{\m\n}\equiv T^{\m\n}+t^{\m\n}$ equals,
\be\label{dtau}
\pa_\m \tau^{\m\n}=\int\left(M{dU^\n\over d\s} -f^\n\right)\delta^4(x-x(\s))\,d\s.
\ee
Requiring  $\tau^{\m\n}$ to be conserved, we {\it derive} then the Lorentz--Dirac equation \eref{ld}.

As on sees, the dynamical information stored in $\pa_\m \tau^{\m\n}$ is supported entirely on the worldline: the relation \eref{dtau} would therefore be completely empty, would we not have been able to construct a $T^{\m\n}$ that is ``defined'' also on the wordline, more precisely, a $T^{\m\n}$ that is a distribution. The structure of the basic identity \eref{divtem} represents, indeed, the core of energy--momentum conservation in a generic particle-- or brane--theory: the exchange of energy--momentum between the brane and the field  occurs precisely {\it on} the brane, while away from the brane the energy--momentum of the field flows freely.

The lesson we learn from this construction is that {\it local} conservation of energy--momentum provides a legitimation of the the self--interaction equation \eref{ld}, that otherwise must be postulated. The rest of the paper represents the first step of the realization of this paradigm for a generic brane in $D$ dimensions, that is, the construction of a renormalized energy--momentum tensor for the field, generalizing \eref{p4}.

\subsection{Self--forces of charged particles and branes}

Before attacking the construction of such a tensor for a generic brane, we summarize briefly -- and without the pretention of completeness --  the situation of the self--interaction forces for particles and branes in $D$ dimensions. As observed above, in most of the cases the techniques for the derivation of these forces involve a regularization of the r.h.s. of  \eref{lor} or, for branes, of \eref{eqb}, aimed to isolate (and subtract) the divergent terms.
This subsection has review character, and its details are not essential for the comprehension of the rest of the paper, since the explicit applications of our approach regard mainly flat branes -- a case where the self--force is zero.

\ul{\it Self--forces for particles.} The self--force equation for a particle in
$D=3$ has been derived in \cite{KLS,G1}, and for a particle in $D=6$ in \cite{KLS,K,Y}. Generalizations for charged particles in higher dimensional {\it even} space--times have been provided  in \cite{MM,G2}. The general feature that arises from these papers is that the  bare self--force $eF^{\m\n}(x(\s))U_\n$ splits in any dimension in a divergent part ${\cal F}^{\m}(\s)$, and in a finite part $f^\m(\s)$,
\be\label{split}
eF^{\m\n}(x(\s))U_\n\quad\rightarrow \quad {\cal F}^\m(\s)+f^\m(\s).
\ee
${\cal F}^{\m}(\s)$ is a sum of  ``local'' terms, i.e. terms that involve only multiple derivatives of the velocity $U^\m(\s)$ at the same instant $\s$. The leading divergent term renormalizes the mass of the particle, in that,
$$
{\cal F}^\m(\s)\bigg|_{lead}\propto {dU^\m\over d\s},
$$
while the renormalization/elimination of the subleading divergences requires the introduction of new interaction parameters in the Lagrangian. ${\cal F}^\m(\s)$ is thus {\it lagrangian}, in the sense that it can be derived from an action. On the other hand, the finite self--force  $f^\m(\s)$ is {\it non--lagrangian}. For {\it even} $D$ it is local and contains as highest derivative a term linear in $\displaystyle{d^{D-2}U^\m\over d\s^{D-2}}$ -- see \eref{fvp4} for $D=4$ -- while for {\it odd} $D$ it is non--local and depends on the entire history of the particle's worldline $x^\m(\s'),\,\forall\,\s'$.

\ul{\it Self--forces for branes.}
For branes the form of the self--force is less well settled, and in general more complicated, but again one has a splitting as in \eref{split}. For a string in $D=4$ a preliminary analysis of the self--interaction
has been performed in \cite{LR,DQ,DH}, while the explicit derivation of the self--force
has been attacked in \cite{CHH}--\cite{BD}. It turns out that in this case ${\cal F}^\m(\s)$ is made out of a single local term, that renormalizes just the string--tension (its mass per unit length),  while $f^\m(\s)$  is non--local. Generic $p$--branes in $D$ dimensions, for which the analysis is much more complicated, have been considered in \cite{CBU,BCM,KS}, with the principal aim of isolating the divergent contribution ${\cal F}^\m(\s)$. To the extent to which the analysis has been performed it emerges that -- as in the case of a particle in $D$ dimensions -- ${\cal F}^\m(\s)$ is made out of a finite number of local terms, while $f^\m(\s)$ is highly non--local and {\it non--lagrangian}. The resulting equations of motion for the brane--coordinates $x^\m(\s)$ are  higher order integro--differential equations, but in the general case explicit formulas for $f^\m(\s)$ and ${\cal F}^\m(\s)$ are missing and/or complicated.

Concluding we may say that the
``unpleasant features'' of the Lorentz--Dirac equation \eref{ld} for a particle in $D=4$, hold true also for the self--interaction equations for branes in $D$ dimensions: a) they can not be {\it derived} from the generalized Maxwell-- and Lorentz--equations \eref{eqa}, \eref{eqb}, and eventually they must be {\it postulated}; b) they are of {\it higher} order in time derivatives and hence in conflict with Newton's determinism; in particular they may change drastically the deterministic picture that arises, instead, when one ignores self--interaction and radiation damping, see e.g. \cite{HK} and references therein; c) they are Lorentz--covariant, but they {\it can not be derived from an action}. Especially this last point raises once more the problem of energy--momentum conservation, since in absence of an action N\"other's theorem can no longer be
enforced.

On the other hand, the ``phenomenological'' relevance of these equations in classical string-- and brane--theory is beyond question.
A safe framework to establish them is therefore required, and the approach we propose is aimed to provide such a universal framework.

\section{Charged $(p-1)$--branes in $D$ dimensions}\label{pbrane}

\subsection{Equations of motion and solutions}

The Electrodynamics of a $(p-1)$--brane in $D$ dimensions, coupled minimally to a rank $p$ antisymmetric potential $A_{\m_1\cdots\,\m_p}$, is governed by the
action, see \eref{fp},
 \be\label{action}
 I={(-)^p\over p!}\int d^Dx \left({1\over 2(p+1)}\, F^{\m_1\cdots\,\m_{p+1}}F_{\m_1\cdots\,\m_{p+1}}
 + A_{\m_1\cdots\,\m_p}\,
 J^{\m_1\cdots\,\m_p}\right)- M\int\sqrt{g}\,d^p\s.
 \ee
$\s^i=(\s^0,\cdots,\s^{p-1})$ are the coordinates one the brane, whose worldvolume is parametrized by the fields $x^\m(\s)$, $\m=0,\cdots,D-1$. We consider a flat target--space with Minkowski metric $\eta^{\m\n}=(1,-1,\cdots,-1)$. The induced metric on the brane is expressed in terms of the tangent vectors, or generalized velocities, $U^\m_i$, $$
g_{ij}= U_i^\m U_j^\n \eta_{\m\n}, \quad\quad U^\m_i\equiv \pa_i
x^\m(\s),
$$
with inverse $g^{ij}$, and $g\equiv(-)^{p+1}\,det \,g_{ij}$. The projectors on the spaces respectively tangent and orthogonal to the brane are then,
 \be\label{proj}
P^{\m\n}= g^{ij}U^\m _i U^\n_j, \quad \quad
Q^{\m\n}=P^{\m\n}-\eta^{\m\n}.
 \ee
Notice that we have the sign--flipped decomposition $\eta^{\m\n}=P^{\m\n}-Q^{\m\n}$, that makes the matrix $Q^{\m\n}$ positive definite. We introduce also the covariant (w.r.t. diffeomorphisms on the brane worldvolume) derivatives of the tangent vectors,
\be\label{covd}
D_i U_j^\m=\pa_iU^\m_j-\Gamma_{ij}^k U^\m_k,
\ee
where $\Gamma_{ij}^k$ is the affine connection associated to $g_{ij}$. The worldvolume indices $i,j$ are raised and lowered with the metric $g_{ij}$.
The current in \eref{action} is given by,
 \be\label{cur}
J^{\m_1\cdots\,\m_p}(x)=e\int d^p\s \,W^{\m_1\cdots\,\m_p}\,
\dl^D(x-x(\s)), \quad\quad  W^{\m_1\cdots\,\m_p}= \ve^{i_1\cdots\,
i_p}\, U^{\m_1}_{i_1}\cdots  U^{\m_p}_{i_p}.
 \ee
$M$ is the tension of the brane, or the mass per unit brane volume (that for $p=1$ reduces to the mass of the particle), and $e$ is its charge. The sign $(-)^p$ in
\eref{action} is chosen such that the energy of the electromagnetic field -- before
renormalization -- is positive definite.

The equations of motion for the gauge field and the brane coordinates
descending from \eref{action} are,
 \ba
 \pa_\m F^{\m\m_1\cdots\, \m_p}&=&J^{\m_1\cdots\,\m_p},\label{eqa}\\
 M D_iU^{\m\, i}   &=& M{1\over \sqrt{g}}\,\pa_i\left(\sqrt{g}\,g^{ij}\,U_j^\m\right)=
 (-)^{p+1}{e\over p!}\,{1\over\sqrt{g}} \,F^{\m\m_1\cdots\,\m_p}\,W_{\m_1\cdots\,\m_p},  \label{eqb}
 \ea
where the electromagnetic field in \eref{eqb} is evaluated
on the brane, i.e. at $x^\m=x^\m(\s)$. The r.h.s. of \eref{eqb} represents the ``bare'' self--force, that is infinite since the field is singular at the brane's position, see below.

The ``bare'' energy--momentum tensor of the electromagnetic field descending from \eref{action} is,
\be\label{th}
\Th^{\m\n}={(-)^p\over p!}\,\left(F^{\m\alpha_1\cdots\,\alpha_p} F^\n{}_{\alpha_1\cdots\,\alpha_p}-{1\over 2(p+1)}\,\eta^{\m\n} F^{\alpha_1\cdots\,\alpha_{p+1}} F_{\alpha_1\cdots\,\alpha_{p+1}}\right),
\ee
while the energy--momentum tensor of the brane is,
\be\label{tmn}
t^{\m\n}=M\int \sqrt{g}\,P^{\m\n}\,\dl^D(x-x(\s))\,d^p\s.
\ee

\ul{\it Solution of Maxwell's equation.}
In the Lorentz--gauge -- $\pa_{\m_1} A^{\m_1\cdots\,\m_p}=0$ -- the generalized Maxwell equation \eref{eqa}  reduces to the d'Alembert--type equation,
\be
\Box A^{\m_1\cdots\,\m_p}=J^{\m_1\cdots\,\m_p}.
\ee
The (retarded) solution of this equation can be written as the convolution,
\be\label{ret}
A^{\m_1\cdots\,\m_p}=G*J^{\m_1\cdots\,\m_p},
\ee
where $G$ is the retarded Green function in $D$ dimensions, satisfying $\Box\, G(x)=\dl^D(x)$. It entails different analytic expressions according to whether $D$ is even or odd \cite{CH},
\be
 G(x)=\cases{\displaystyle
 {H(x^0)\over 2\pi^{N+1}} \, \left({d\over dx^2}\right)^N\dl(x^2),  & for $
D=2N+4$, \cr
 &\cr\label{green}
 \displaystyle{H(x^0)\over 2\pi^{N+1}} \,\left({d\over dx^2}\right)^N{H(x^2)\over \sqrt{x^2}}, & for $D=2N+3$,\cr}
 \ee
where $x^2=x^\m x_\m$.

\ul{\it Analysis of the singularities.} We discuss now qualitatively the singularities of the field and the energy--momentum tensor in the vicinity of the brane, implied by  \eref{ret} and \eref{green}. Indicate the coordinates normal to the brane (see below for a precise definition) with $r^a$, $a=1,\cdots, n$, where the ``codimension'' $n$ is defined by,
\be\label{dp}
n=D-p.
\ee
In particular, at the brane, where $x^\m=x^\m(\s)$, by definition we have  $r^a=0$. The leading singular behavior of the field \eref{fp} for $r^a\ra 0$ is then schematically, see section \ref{general},
 \be\label{beha}
F^{\m_1\cdots\,\m_{p+1}}\sim {1\over r^{n-1}}, \quad r=\sqrt{r^ar^a}.
 \ee
At the brane the field diverges thus, giving rise to the aforementioned  infinite self--force. On the other hand,
due to \eref{beha} also the energy--momentum tensor \eref{th} diverges near the brane as,
\be\label{tdiv2}
\Th^{\m\n}\sim {1\over r^{2n-2}},
\ee
relation that generalizes \eref{tdiv}.
The singular behavior \eref{tdiv2} causes the pathologies illustrated in subsection \ref{udf} in the case of a particle in $D=4$. More concretely, if $n\ge2$ the energy--momentum integrals,
 \be\label{ptot}
P^\m = \int \Th^{0\m}\,d^{D-1}x\sim\int \Th^{0\m}\,d^nr\, d^{p-1}\s\sim \int {dr\over r^{n-1}}\,\,d^{p-1}\s,
 \ee
diverge near the brane, where $r\ra 0$ \footnote{Here we are not dealing with the infrared
divergences that may appear in the total $P^\m$, if the
brane is infinitely extended.}. Stated differently, due to \eref{tdiv2}, if $n\ge 2$ the functions $\Th^{\m\n}$ are not distributions, since they are not locally integrable in $D$ dimensions. Consequently also the $D$--divergence $\pa_\m \Th^{\m\n}$ is ill--defined.

For $n=0,1$, instead, the self--force in \eref{eqb} is finite and the bare energy--momentum tensor in \eref{th} is a distribution. In this paper we will always consider a codimension $n\ge2$.

\subsection{Renormalized energy--momentum tensor: general construction}\label{rem}

In this subsection we present our general recipe to overcome the problems just mentioned. As in the case of a particle in $D=4$ it consists of a regularization, followed by the subtraction of local counterterms.

The recipe we propose is rather simple. We introduce in $D$ space--time dimensions a regularized Green function $G_\ve(x)$ --  replacing \eref{green} -- that preserves $D$--dimensional Lorentz--invariance, as well as $p$--dimensional diffeomorphism invariance (for previous applications see \cite{LR,KS,Da}),
\be
 G_\ve(x)=\cases{\displaystyle
 {H(x^0)\over 2\pi^{N+1}} \,\left(-{d\over d\ve^2}\right)^N\dl(x^2-\ve^2),& for $
D=2N+4$, \cr
 &\cr\label{greene}
 \displaystyle{H(x^0)\over 2\pi^{N+1}} \,\left(-{d\over d\ve^2}\right)^N{H(x^2-\ve^2)\over \sqrt{x^2-\ve^2}}, & for $D=2N+3$.\cr}
 \ee
Notice that in these formulae, that generalize the regularized kernel \eref{ge4} in  $D=4$, the derivative $-d/d\ve^2$ can equally well be replaced by $d/dx^2$. In particular we have,
$$
{\rm Lim}\,_{\ve\ra 0}\,G_\ve=G.
$$
Recall that with the capital ``${\rm Lim}$'' we mean always the limit in the sense of distributions. We introduce then regularized potentials and fields via, compare with \eref{amue},
\be\label{rete}
A^{\m_1\cdots\,\m_p}_\ve=G_\ve*J^{\m_1\cdots\,\m_p}, \quad\quad F_\ve^{\m_1\cdots\,\m_{p+1}}= (p+1)\,\pa^{[\m_1} A_\ve^{\m_2\cdots\,\m_{p+1}]},
\ee
that are indeed regular on the brane, i.e. at $x^\m=x^\mu(\s)$. In particular,  \eref{cur} and \eref{rete} give,
\be\label{ae}
A^{\m_1\cdots\,\m_p}_\ve(x)=e\int \sqrt{g}\,\,W^{\m_1\cdots\,\m_p}(\s)\,G_\ve(x-x(\s))\,d^p\s.
\ee
Eventually we define the regularized energy--momentum tensor, replacing \eref{th},
\be\label{the}
\Th^{\m\n}_\ve={(-)^p\over p!}\,\left(F_\ve^{\m\alpha_1\cdots\,\alpha_p} F_\ve{}^\n{}_{\alpha_1\cdots\,\alpha_p}-{1\over 2(p+1)}\,\eta^{\m\n} F_\ve^{\alpha_1\cdots\,\alpha_{p+1}} F_{\ve\,\alpha_1\cdots\,\alpha_{p+1}}\right).
\ee
For $\ve>0$ the functions \eref{the} are now regular on the brane -- actually, they are $C^\infty$--functions -- and in particular they are {\it distributions}. We will illustrate these features explicitly for a generic flat brane in section \ref{stat}.

\subsubsection{Divergent and finite counterterms}\label{daf}

We formulate now our proposal for the renormalized energy--momentum tensor $T^{\m\n}$ for the field of a generic brane, generalizing \eref{p4}. We set,
\ba\label{tren}
T^{\m\n}&=&\wt T^{\m\n}+D^{\m\n}\\
\wt T^{\m\n}&=&{\rm Lim}\,_{\ve\ra 0}\left(\Th^{\m\n}_\ve-\wh\Th_\ve^{\m\n}\right),\label{ttilde}
\ea
where the ``divergent'' and ``finite'' counterterms $\wh\Th_\ve^{\m\n}$ and $D^{\m\n}$  are respectively of the form,
\ba
\wh\Th_\ve^{\m\n}&=&\int\sqrt{g}\,\,R^{\m\n}_\ve\,\dl^D(x-x(\s))\,d^p\s,\label{re}\\
D^{\m\n}&=&\int\sqrt{g}\,\,\D^{\m\n}\,\dl^D(x-x(\s))\,d^p\s.\label{dmn}
\ea
Both counterterms are local, where with ``local'' here and in the following  we mean ``supported on the brane''.
$R^{\m\n}_\ve$ and $\D^{\m\n}$ are symmetric tensors living on the brane worldvolume. $R^{\m\n}_\ve$ is a tensor that depends on $\ve$ and diverges as $\ve\ra 0$ (in the ordinary sense), while $\D^{\m\n}$ is independent of $\ve$.

$R^{\m\n}_\ve$ has to be chosen in such a way that the ${\rm Lim}\,_{\ve\ra0}$ in \eref{ttilde} exists. The local form of \eref{re} is suggested by the fact that  $\Th^{\m\n}$ is singular on the brane, but regular in the complement of the brane. The counterterm $\wh\Th_\ve^{\m\n}$ represents thus just the divergent part of $\Th^{\m\n}_\ve$, as $\ve\ra 0$ in the sense of distributions.
As we will see in subsection \ref{tdc}, the leading divergences of $R^{\m\n}_\ve$ for $\ve\ra 0$ are,
\be
R^{\m\n}_\ve\sim \cases
{
\displaystyle{1\over \ve^{n-2}}, &for $n>2$,
\cr
&\cr
\ln\ve, &for $n=2$.
\cr
}
\ee
Thanks to the fact that our regularization  preserves
$D$--dimensional Lorentz invariance and $p$--dimensional diffeomorphisms, $R^{\m\n}_\ve$ is necessarily a tensor under these symmetries.

Whereas
$\wh\Th_\ve^{\m\n}$ realizes a ``minimal subtraction'' -- including only divergent terms -- $D^{\m\n}$ represents a {\it finite} local counterterm, that must be a tensor, too. No such term is present in \eref{p4} -- that corresponds to a case with codimension $n=3$ -- but we will see that for $n$ {\it even} $D^{\m\n}$ is necessarily non--vanishing. The reason for this has to do with the consistent definition of the self--force, as will be explained in the next paragraph.

In conclusion, if there exists an $R^{\m\n}_\ve$ such that the limit \eref{ttilde} exists, the resulting $T^{\m\n}$ is a {\it distribution--valued tensor}. Moreover, since the counterterms  \eref{re}, \eref{dmn} are both supported on the brane,  in the complement of the brane $T^{\m\n}$ coincides with $\Th^{\m\n}$.

\subsubsection{Derivation of the self--force and energy--momentum conservation}\label{dot}

Once the existence of the distributional limit \eref{ttilde} has been established, the existence of the distributional divergence $\pa_\m T^{\m\n}$ is guaranteed.  Moreover, it is necessarily of the form,
\be\label{selff}
\pa_\m T^{\m\n}=-\int\sqrt{g}\,f^\n \,\dl^D(x-x(\s))\,d^p\s,
\ee
for some vector $f^\n$ (that could involve also derivative operators acting on $\dl^D(x-x(\s))$, see below). This is due to the fact that, thanks to  \eref{eqa}, the ``naive'' divergence of the bare tensor \eref{th} turns out to be,
$$
\pa_\m \Th^{\m\n}={(-)^p\,e\over p!}\,\int F^{\n\a_1\cdots\,\a_p} \,W_{\a_1\cdots\,\a_p} \,\dl^D(x-x(\s))\,d^p\s.
$$
This shows that in the complement of the brane $\pa_\m \Th^{\m\n}$ vanishes. Since, on the other hand, in the complement of the brane $\Th^{\m\n}_\ve$ converges {\it pointwise} to $\Th^{\m\n}$, and since also the counterterms \eref{re}, \eref{dmn} are supported on the brane, it follows that  $\pa_\m T^{\m\n}$ vanishes in the complement of the brane. This means that $\pa_\m T^{\m\n}$ is necessarily supported on the brane, and hence of the form  \eref{selff}.

Keeping for the energy--momentum tensor of the brane the expression \eref{tmn}, that satisfies,
\be\label{dtmn0}
\pa_\m t^{\m\n}=M\int \sqrt{g}\,\,D_iU^{\n\, i}\,\dl^D(x-x(\s))\,d^p\s,
\ee
and defining the total energy--momentum tensor again as $\tau^{\m\n}=T^{\m\n}+t^{\m\n}$,
with the aid of \eref{selff} we obtain,
\be
\pa_\m \tau^{\m\n}=\int\sqrt{g}  \left[M D_iU^{\n\,i}- f^\n\right]\dl^D(x-x(\s))\,d^p\s.
\ee
If the total energy--momentum has to be conserved we are thus tempted to impose on the brane coordinates $x^\m(\s)$ the generalized Lorentz--Dirac equation,
\be\label{self}
M D_iU^{\m\,i}=f^\m.
\ee
This equation replaces thus the ill--defined Lorentz equation \eref{eqb},
and consequently the vector $f^\m$ represents the, now finite, self--force.

\subsubsection{Finite counterterms and consistency conditions}\label{cc}

The construction just given is subjected to two consistency conditions that the self--force $f^\m$ must fulfill, due to the structure of equation \eref{self}.

\ul{\it Self--force orthogonal to the brane.} The first consistency condition derives from the fact that the l.h.s. of \eref{self} satisfies identically the constraint,
\be\label{udu}
U_{\m j}\,D_iU^{\m\,i}=0.
\ee
This follows from the identities,
$$
U^{\m\,i}D_jU_{\m\,i}=0,\quad\quad D_iU_j^\m=D_jU_i^\m.
$$
As a consequence the self--force must be orthogonal to the brane,
\be\label{orto}
U_\m^if^\m=0.
\ee
If \eref{self} would be derivable from an action -- but we know that this is not true -- then \eref{orto} would be implied automatically by diffeomorphism invariance. In the present case the validity of \eref{orto} is, instead, not guaranteed, and it must be checked a posteriori. Notice that the Lorentz--Dirac self--force \eref{fvp4} does indeed satisfy this relation.

\ul{\it Algebraic and operatorial self--forces: finite counterterms.} The second consistency condition arises from the fact that the vector $f^\m$ identified in \eref{selff} can be ``algebraic'', i.e. correspond to a function $f^\m(\s)$ on the worldvolume, or ``operatorial'', i.e. involve also space--time derivatives. An example of the first kind is given by the Lorentz--Dirac force \eref{fvp4}, $f^\m\propto{d^2U^\m\over d\s^2}+\left({dU\over d\s}\right)^2U^\m$, and an example of the second kind is $f^\m\propto \pa^\m$. Covariance reasons allow indeed for both types of forces in \eref{selff}. However, if $f^\m$ is operatorial, equation \eref{self} would not make sense, and consequently $\tau^{\m\n}$ would not be conserved, {\it whatever equation of motion one imposes on $x^\m(\s)$} !

It is at this point that the finite counterterm $D^{\m\n}$ in \eref{tren} comes into the game. The starting point of our approach is indeed the tensor $\wt T^{\m\n}$ in \eref{ttilde}, that arises from a ``minimal subtraction''. Evaluating its divergence we obtain,
\be\label{dwtmn}
\pa_\m \wt T^{\m\n}= -\int\sqrt{g}\,\wt f^\n \,\dl^D(x-x(\s))\,d^p\s,
\ee
for some vector $\wt f^\n$. It $\wt f^\n$ is algebraic, we set $D^{\m\n}=0$, $T^{\m\n}=\wt T^{\m\n}$, giving $ f^\n=\wt f^\n$, and \eref{self} is a well--posed equation. If, on the contrary, $\wt f^\n$ is operatorial, we take $D^{\m\n}\neq0$ and \eref{tren} gives then,
\be\label{algf}
\pa_\m T^{\m\n}=-\int\sqrt{g}\,\left(\wt f^\n-\D^{\m\n}\pa_\m\right)\dl^D(x-x(\s))\,d^p\s.
\ee
We can therefore search for a tensor $\D^{\m\n}$ such that the vector,
\be\label{fnt}
f^\n\equiv  \wt f^\n-\D^{\m\n}\pa_\m,
\ee
becomes algebraic \footnote{The vectors $f^\n$ and $\wt f^\n$ are defined modulo terms that become total derivatives when inserted in \eref{algf}, since they drop then out from the integral.}. In this case equation \eref{self} can be imposed consistently, and the vector $f^\m$ in \eref{fnt} identifies then the self--force of the brane. In conclusion, the second consistency condition -- arising from the requirement of energy--momentum conservation -- is that there exists a tensor $\D^{\m\n}$ such that \eref{fnt} becomes algebraic.

Local finite counterterms of the form \eref{dmn} -- with a completely identical role, i.e. that of restoring conservation of the energy--momentum tensor -- have been employed also in \cite{Ro} for a particle in $D=4$. In our approach, for a particle in $D=4$, corresponding to  $n=3$, these terms are actually absent, see  \eref{p4}. On the contrary, as mentioned above, such finite counterterms will show up for every {\it even} $n$, see section \ref{general}.

Our approach for the construction of a renormalized energy--momentum tensor is reminiscent of a procedure necessary in quantum field theory, when the regularization breaks a local symmetry (like chiral gauge symmetry, or local conformal invariance). In that case after  subtraction of the {\it divergent} local counterterms from the effective action, a priori one must  subtract also {\it finite} local counterterms, to restore the symmetry. Only if no such finite counterterms exist the theory is, actually, ``anomalous''. In the present case the subtraction of finite local counterterms is needed to restore the conservation of the energy--momentum tensor, rather than a symmetry.

\subsubsection{Uniqueness of $T^{\m\n}$ and $f^\m$ }\label{uni}

The construction presented above involves the finite counterterm \eref{dmn} that -- by the very definition of a ``finite counterterm'' -- is a priori not uniquely determined.
Nevertheless, we show now that our approach, if successful, leads to a uniquely defined $T^{\m\n}$, and hence also to a unique self--force.
To this end we remember that on physical grounds the energy--momentum tensor is in general defined modulo a $D$--divergence,
\be\label{ambi}
T'^{\m\n} = T^{\m\n} +\pa_\rho V^{\rho\m\n},\quad\quad \mbox{with}\quad
V^{\rho\m\n}=-V^{\m\rho\n},
\ee
in that $T'^{\m\n}$ entails the same total energy--momentum integrals as $T^{\m\n}$. To keep $T'^{\m\n}$ symmetric we require in addition,
$$
\pa_\rho V^{\rho\m\n}=\pa_\rho V^{\rho\n\m}.
$$
In particular, since $\pa_\m T'^{\m\n}=\pa_\m T^{\m\n}$
the modification \eref{ambi} leaves the self--force unchanged.

Consider now a generic change of the finite counterterm in \eref{dmn},  $\Delta^{\m\n}\ra \Delta^{\m\n}+l^{\m\n}$. This replacement is consistent if the tensor,
$$
L^{\m\n}=\int\sqrt{g}\,\,l^{\m\n}\,\dl^D(x-x(\s))\,d^p\s,
$$
satisfies,
\be\label{damn}
\pa_\m L^{\m\n}=\int\sqrt{g}\,\,l^{\m\n}\,\pa_\m\dl^D(x-x(\s))\,d^p\s=-
\int\sqrt{g}\,b^\n\,\dl^D(x-x(\s))\,d^p\s,
\ee
for some {\it algebraic} vector $b^\m$ obeying,
\be\label{ortob}
U_\m^i\,b^\m=0,
\ee
see \eref{orto}.
In fact, in this case the energy--momentum tensor and the self--force would change as follows,
\be\label{shift}
T^{\m\n}\ra T^{\m\n}+L^{\m\n}, \quad\quad f^\m\ra f^\m+b^\m.
\ee
We analyze now the possible structures of the tensors $l^{\m\n}$ and show that they lead, 1) to a redefinition of the brane tension or, 2) to modifications of the type \eref{ambi}. On general grounds $l^{\m\n}$ is a linear combination of operators of the type,
$$
l^{\m\n}= C^{\m\n\a_1\cdots\,\a_N}(\s)\,\pa_{\a_1}\cdots\pa_{\a_N},
$$
where $C^{\m\n\a_1\cdots\,\a_N}(\s)$ are tensor fields on the brane
\footnote{In this proof we are tacitly assuming that the coefficients $C^{\m\n\a_1\cdots\,\a_N}(\s)$ are polynomials of $U^\m_i(\s)$ and their (multiple) covariant derivatives.}.  We distinguish two cases.

\ul{\it $l^{\m\n}$ without derivatives.} If $l^{\m\n}$ does not contain derivatives it has the general form,
\be\label{lmn0}
l^{\m\n}=\Phi_1\, P^{\m\n}+ \Phi_2\,\eta^{\m\n},
\ee
where $\Phi_i(\s)$ are scalar fields on the brane.
The condition \eref{damn} gives then,
$$
b^\m= -\pa_i\Phi_1\, U^{\m\,i} -\Phi_1 \,D_iU^{\m\,i}- \Phi_2\, \pa^\m.
$$
Since we want $b^\m$ to be algebraic, $\Phi_2$ must vanish.
\eref{ortob} implies then that $\Phi_1$ is constant, see \eref{udu}, and hence,
$$
b^\m=-\Phi_1 D_iU^{\m\,i}.
$$
The shift of the self--force \eref{shift} corresponds thus just to the redefinition of the brane tension $M\ra M+\Phi_1$, see \eref{self}. The modification  \eref{lmn0} is therefore physically irrelevant.

\ul{\it $l^{\m\n}$ with derivatives.} If $l^{\m\n}$ contains derivatives, it must contain at least two of them. The relation \eref{damn} produces then necessarily an {\it operatorial} $b^\m$, unless $b^\m$ {\it vanishes}. This means that we must have $\pa_\m L^{\m\n}=0$ {\it identically}. Due to algebraic reasons there exists then always a tensor $V^{\rho\m\n}$, with $V^{\rho\m\n}=-V^{\m\rho\n}$, such that $L^{\m\n}=\pa_\rho V^{\rho\m\n}$. Instead of giving the general proof, we illustrate the situation with two examples:
\ba\label{l1}
L^{\m\n}_1&=&\int\sqrt{g}\, C_1\left(\eta^{\m\n}\Box-\pa^\m\pa^\n\right)\dl^D(x-x(\s))\,d^p\s,\\
L^{\m\n}_2&=&\int\sqrt{g}\,C_2\left(P^{\m\n}\Box-2\,P^{\a(\m}\,\pa^{\n)}\pa_\a+\eta^{\m\n}P^{\a\bt}
\pa_\a\pa_\bt\right)\dl^D(x-x(\s))\,d^p\s.\label{l2}
\ea
Here the $C_i$ are generic linear combinations of operators of the type $C^{\a_1\cdots\,\a_N}(\s)\,\pa_{\a_1}\cdots\pa_{\a_N}$.  By inspection one has $\pa_\m L_i^{\m\n}=0$, and it is easily seen that $L^{\m\n}_i=\pa_\rho V^{\rho\m\n}_i$, $V^{\rho\m\n}_i=-V^{\m\rho\n}_i$, where,
\ba
V^{\rho\m\n}_1&=&\int\sqrt{g}\,C_1\left(\eta^{\m\n}\pa^\rho-\eta^{\rho\n}\pa^\m\right)
\dl^D(x-x(\s))\,d^p\s,\nn\\
V^{\rho\m\n}_2&=&\int\sqrt{g}\,C_2\left(P^{\m\n}\pa^\rho-P^{\rho\n}\pa^\m+\eta^{\m\n}P^{\rho\a}\pa_\a -\eta^{\rho\nu}P^{\m\a}\pa_\a\right)\dl^D(x-x(\s))\,d^p\s.\nn
\ea
Tensors $l^{\m\n}$ containing derivatives give thus rise to physically irrelevant modifications of $T^{\m\n}$ of the type \eref{ambi}, and they do therefore not modify the self--force.

In conclusion, our approach leads to {\it uniquely determined energy--momentum tensors and self--forces}.

\subsubsection{Finite energy--momentum integrals}\label{fem}

Once $T^{\m\n}$ is a distribution, it entails automatically {\it finite} energy--momentum  integrals  $P^\m_V(t)$ over any bounded volume $V$, irrespective of the fact that $V$ encloses (a portion of) the brane or not.

To define them we introduce the characteristic function of the volume $\chi_V(\vec x)$, and regard it as a test--function \footnote{The function $\chi_V$ is not a ``true'' test function belonging to ${\cal S}(\mathbb{R}^{D-1})$, since it is discontinuous at the boundary of $V$. Therefore, from a rigorous point of view the quantities \eref{pmom} should rather be written as the limits $\lim_{\a\ra 0}T^{0\m}\left(\chi_V^\a\right)$, where $\chi_V^\a$ are smooth differentiable approximations of $\chi_V$. However, as long as at the instant $t$ the brane is not tangent to the boundary of $V$, this limiting process is not necessary.} in  $\mathbb{R}^{D-1}$. Similarly we consider $T^{0\m}(t,\vec x)$ in \eref{tren} -- at fixed time $t$ -- as a distribution in ${\cal S}'(\mathbb{R}^{D-1})$. The energy--momentum in $V$ at time $t$ can then be defined
as the distribution $T^{0\m}(t,\vec x)$ applied to the test function $\chi_V$,
\ba
P^\m_V(t)&=&T^{0\m}\left(\chi_V\right)=
\lim_{\ve\ra 0}\left[\Th^{0\m}_\ve(\chi_V) -\wh\Th_\ve^{0\m}(\chi_V)\right]
+D^{0\m}(\chi_V)\nn\\
&=&\lim_{\ve\ra0}\int_V\left(\Th^{0\m}_\ve-\wh\Th_\ve^{0\m}+D^{0\m}\right)d^{D-1}x.
\label{pmom}
\ea
As long as the brane lies outside $V$, the counterterms $\wh\Th_\ve^{0\m}$ and $D^{0\m}$ do not contribute to the integral since they are supported at the brane's position, and one gets back the ``bare'' finite momentum integrals: $P^\mu_V = \lim_{\ve\ra 0}\int_V  \Th^{0\m}_\ve \, d^{D-1}x= \int_V \Th^{0\m} \, d^{D-1}x.$ However, if the volume $V$ encloses (a portion of) the brane, the subtraction of $\wh\Th_\ve^{0\m}$ becomes crucial to make the limit \eref{pmom} finite. On the other hand, the addition of $D^{0\m}$ is crucial to make the energy--momentum conserved.

\section{Flat branes}\label{stat}

In the following sections we exemplify the approach presented above in the case of flat branes. In this section we derive thus the basic properties of the (regularized) dynamics of these branes, the main result being an explicit expression of the regularized energy--momentum tensor for the field of a generic flat $(p-1)$--brane in $D$ dimensions, see \eref{theg}.

A ``flat brane'' represents in some sense a generalization of a particle in uniform motion. By definition such a brane fulfills the covariant constraint,
\be\label{statp}
D_i U_j^\m=\pa_iU^\m_j-\Gamma_{ij}^k U^\m_k=0,
\ee
meaning that the generalized velocities are covariantly constant. Consequently
the manifold described by the brane during its evolution is flat. In fact, applying a second derivative to \eref{statp} and antisymmetrizing the indices, one deduces that the Riemann tensor vanishes,
$$
0=(D_kD_i-D_iD_k)U_j^\m= R_{kij}{}^lU^\m_l\quad\Rightarrow\quad  R_{kijl}=0.
$$

\ul{\it Energy--momentum conservation and vanishing self--force.} Since \eref{statp} holds for all $\m,i,j$, from \eref{dtmn0} we see that the energy--momentum tensor of the brane is separately conserved,
$$
\pa_\m t^{\m\n}=0,
$$
and from \eref{self} we see that the self--force must be zero,
$$
f^\m=0.
$$
From \eref{selff} we conclude then that for a flat brane also the renormalized energy--momentum tensor of the field must be separately conserved,
\be\label{sep}
\pa_\m T^{\m\n}=0.
\ee

\ul{\it Flat coordinates.} Since the Riemann tensor is zero, through a diffeomorphism on the brane we can go over to ``flat'' coordinates $\s^i$, in which the induced metric becomes flat and minkowskian, at least locally, $g_{ij}=\eta_{ij}$, $diag(\eta_{ij})=(1,-1,\cdots,-1)$, $\G_{ij}^k=0$. In these coordinates the tangent vectors are then constant,
\be\label{coord}
\pa_iU^\m_j=0\quad\Rightarrow\quad U^\m_j= {\mbox constant}\quad\Rightarrow \quad x^\m(\s)=U^\m_i\s^i+x^\m(0).
\ee
In particular we have then,
$$
U_i^\m U_j^\n \eta_{\m\n}= \eta_{ij},\quad \sqrt{g}=1.
$$
In the following we will always use flat coordinates and set, without loss of generality, $x^\m(0)=0$. Since in flat coordinates all tangent vectors are constant, in such coordinates flat branes appear rigid and in uniform motion.

\ul{\it Static coordinates.} Sometimes we will step from flat coordinates to ``static'' coordinates, that can always be reached through a target--space Lorentz transformation. These coordinates are constructed arranging the $x^\m$ into ``parallel'' and ``normal'' coordinates,
\be\label{stat1}
x^\m=(y^i,r^a),\quad i=0,1,\cdots,p-1,\quad\quad a=1,2,\cdots,n,\quad\quad n=D-p.
\ee
By definition in static coordinates the parametrization \eref{coord} simplifies to,
\ba
y^i(\s)&=&\s^i,\label{stat2}\\
r^a(\s)&=& 0,\label{stat3}
\ea
such that the tangent vectors become,
\be\label{stat4}
U_i^j=\dl^j_i,\quad\quad U_i^a=0.
\ee
In these coordinates the brane fields $x^\m(\s)$  for $\m\neq 0$ are time--independent, and the brane appears thus as ``static''.

Since $Q^{\m\n}$ is the orthogonal projector,
in static coordinates we have in particular,
\be\label{Qstat}
Q_{\m\n}\,x^\m x^\n= \eta^{ij}U^\m_i U^\n_jx_\m x_\n-  x^\m x_\m = r^ar^a\equiv r^2.
\ee
In the following we will denote the $\dl$--function on a generic brane with,
\be\label{dn}
\dl^n\equiv\int\sqrt{g}\,\,\dl^D(x-x(\s))\,d^p\s.
\ee
For a flat brane in static coordinates we have then simply,
\be\label{dns}
\dl^n=\dl^n(\vec r).
\ee
Notice that we are allowed to use freely flat and/or static coordinates, since our approach for the construction of a consistent energy--momentum tensor preserves $p$--diffeomorphism as well as $D$--Lorentz invariance.

\ul{\it Regularized potential.} In flat coordinates, that we use from now on, \eref{ae} becomes,
\be\label{ae2}
A^{\m_1\cdots\,\m_p}_\ve(x)=e \,W^{\m_1\cdots\,\m_p}\int G_\ve(x-U\cdot\s)\,d^p\s,\quad\quad x-U\cdot\s\equiv x^\m-U^\m_i\s^i,
\ee
since the tangent vectors are constant. The evaluation of the integral in \eref{ae2} requires different strategies for even and odd $D$, due to the different expressions of the Green functions \eref{greene}; nevertheless, the potentials $A^{\m_1\cdots\,\m_p}_\ve$ have the same analytical expression. The explicit calculation is performed most easily in static coordinates -- see appendix A -- and the result is,
\be
\label{aen}
A^{\m_1\cdots\,\m_p}_\ve(x)=\cases
{
\displaystyle
  {e\,W^{\m_1\cdots\,\m_p}\over 4\pi^{n/2}}\,{\Gamma\left({n\over 2}-1\right)\over\left(Q_{\m\n}\,x^\m x^\n +\ve^2\right)^{n/2-1}},  &  for $n>2$,
\cr
&\cr
&\cr
\displaystyle
-{e\,W^{\m_1\cdots\,\m_p}\over 4\pi}\,\ln\left(Q_{\m\n}\,x^\m x^\n +\ve^2\right),  & for $n=2$.
\cr
}
\ee
Notice that, thanks to \eref{Qstat}, in static coordinates the regularized potential $A^{\m_1\cdots\,\m_p}_\ve(x)$ depends only on the normal coordinates $r^a$, as does the bare one $A^{\m_1\cdots\,\m_p}(x)$. Note also that the expressions \eref{aen} depend not separately on $D$ and $p$, but only on the codimension $n=D-p$. The codimension $n=2$ plays a peculiar role, due to an infrared divergence originating from the infinite extension of a flat brane. As a consequence the potential for $n=2$ -- as given in \eref{ae2} -- would be, actually, divergent. The expression given in  \eref{aen} emerges, indeed, after subtraction of an infrared divergent {\it constant} term: the electromagnetic {\it field} is thus in any case finite.

\ul{\it Regularized field.} The regularized field descending from \eref{aen} and \eref{rete} is,
\be\label{fe}
F^{\m_1\cdots\,\m_{p+1}}_\ve(x)=
-{e\,(p+1) \over \Omega_n}\,{W^{[\m_1\cdots\,\m_p}\,Q^{\m_{p+1}]\n}\,x_\n\over \left(Q_{\m\n}\,x^\m x^\n +\ve^2\right)^{n/2}},
\ee
where we introduced the solid angle in $n$ dimensions,
$$
\Omega_n={2\pi^{n/2}\over \G\left({n\over2}\right)}.
$$
The expression \eref{fe} is now valid also for $n=2$. Since in static coordinates $Q_{\m\n}\,x^\m x^\n=r^2$, for $\ve=0$ the singular behavior of the field near the brane is indeed as in \eref{beha}. On the contrary, for $\ve\neq 0$ the expression \eref{fe} is everywhere finite, also on the brane where $r^a=0$.

\ul{\it Regularized energy--momentum tensor.} Eventually we can now write explicitly the regularized energy--momentum tensor of the  field of a generic flat brane. Inserting \eref{fe} into \eref{the}, and recalling the definition of the $W$--tensor in \eref{cur}, it is straightforward to obtain,
\be\label{theg}
\Th^{\m\n}_\ve(x)=-\left({e\over \Omega_n}\right)^2{Q^{\m\a}\,Q^{\n\bt}\,x_\a x_\bt +\left({1\over 2}\,\eta^{\m\n}-P^{\m\n}\right)Q_{\a\bt}x^\a x^\bt\over \left(Q_{\a\bt}\,x^\a x^\bt +\ve^2\right)^n},
\ee
where the projectors $Q$ and $P$ are given in \eref{proj}. For $\ve\ra0$ this tensor tends pointwise to the bare energy--momentum tensor -- that is not a distribution and behaves near the brane as anticipated in \eref{tdiv2}. On the contrary, the regularized tensor \eref{theg} is regular on the brane. In fact, for a point on the brane, say for $x^{*\m}=U^\m_i\s^i$, we have,
$$
Q^{\a\bt}x^*_\bt=(U^\a_i U^\bt_j\eta^{ij}-\eta^{\a\bt})U_{\bt k}\,\s^k=0.
$$
For $x^\m=x^{*\m}$ the numerator of \eref{theg} is thus zero, while the denominator remains different from zero, reducing to $\ve^{2n}$. The value of the regularized energy--momentum tensor on the brane is therefore finite: $\Th^{\m\n}_\ve(x^*)=0$.

By inspection the basic expression \eref{theg} is, actually, a {\it distribution} that is a $C^\infty$--function, and it marks the starting point of our approach, see \eref{ttilde}.

\ul{\it Conservation.}
For later use we compute the divergence of \eref{theg}. Since $\Th^{\m\n}_\ve$ is a $C^\infty$--distribution, its derivatives can be computed simply ``in the sense of functions'',
\be\label{divteg}
\pa_\m\Th^{\m\n}_\ve=\left({e\over \Omega_n}\right)^2{n\,\ve^2\,Q^{\n\m}\,x_\m \over \left(Q_{\a\bt}x^\a x^\bt+\ve^2\right)^{n+1}}=-\left({e\over \Omega_n}\right)^2 \pa^\n {\ve^2 \over 2\left(Q_{\a\bt}x^\a x^\bt+\ve^2\right)^n} \neq 0.
\ee
Our regularization violates thus the conservation of the energy--momentum tensor, anticipating the potential occurrence of finite counterterms. Notice,
however, that the r.h.s. of \eref{divteg} is proportional to $\ve^2$. This means that in the complement of the brane, i.e. for $Q_{\m\n}\,x^\m x^\n\neq 0$, for $\ve\ra 0$ $\pa_\m\Th^{\m\n}_\ve$ converges {\it pointwise} to zero, as expected. However, for $n>2$ the distributional limit ${\rm Lim}_{\,\ve\ra0}\,\pa_\m\Th^{\m\n}_\ve$ does not exist. More precisely, in the sense of distributions one has the leading behaviors, see subsection \ref{consg},
\be
\pa_\m\Th^{\m\n}_\ve\sim \cases
{\displaystyle
{1\over \ve^{n-2}}, &for $n>2$,
\cr
&\cr
o(1), &for $n=2$.
\cr
}
\ee
It is only after subtraction of $\wh\Th_\ve^{\m\n}$  -- and addition of $D^{\m\n}$ -- that one can regain a conserved energy--momentum tensor.

\section{Particle in $D=6$}\label{p6}

We implement now our approach for the construction of a consistent energy--momentum tensor for a point--particle in $D=6$. In practice this amounts to an explicit realization of  formulae  \eref{tren}--\eref{dmn}. A particle in $D=6$ corresponds to the {\it odd} codimension $n=5$, and this case allows to illustrate some of the basic features of our approach. If the codimension is instead {\it even}, additional features will show up, basically the necessity of finite  counterterms, and these will be illustrated in section \ref{s4} for a string in $D=4$, corresponding to $n=2$.

\subsection{Construction of the divergent counterterm}

The determination of the divergent counterterm \eref{re} is most easily performed using static coordinates $x^\m=(y^i,r^a)$, because then the regularized tensor \eref{theg} depends only on the normal coordinates $r^a$; in the present case $a=1,\cdots,5$, and $y^i\equiv y^0=x^0$. The main result of this subsection is the explicit expression of the counterterm \eref{cou}.

We begin by writing explicitly the components of \eref{theg} in static coordinates,
\ba
\Th_\ve^{00}&=& \left({e\over \Omega_5}\right)^2{r^2\over 2\, (r^2+\ve^2)^5},\label{t1}\\
\Th_\ve^{0a}&=& 0,\\
\Th_\ve^{ab}&=& \left({e\over \Omega_5}\right)^2{r^2\,\dl^{ab}-2\,r^ar^b\over 2\,(r^2+\ve^2)^5},\label{t3}
\ea
with $\Omega_5=8\pi^2/3$.
To find the counterterm we must isolate from $\Th_\ve^{\m\n}$ the terms that diverge as $\ve\ra 0$ in the sense of distributions \footnote{To be precise we work in the space ${\cal S}'(\mathbb{R}^D)$ of {\it tempered} distributions, and hence $\vp(x)$ must  belong to ${\cal S}(\mathbb{R}^D)$, i.e. the space of $C^\infty$--functions on $\mathbb{R}^D$ that vanish for $x^\mu\ra \infty$, together with all their derivatives, more rapidly as any inverse power of the coordinates.}. To this order we must apply $\Th_\ve^{\m\n}$ to a test function $\vp(x)$ and isolate the terms that diverge, in the ordinary sense, as $\ve\ra 0$. In the present case it is sufficient to consider $\vp$ as a function of only the normal coordinates $r^a$, as $\Th_\ve^{\m\n}$ is independent of $y^i$.

To find the divergent parts of \eref{t1}--\eref{t3} it is sufficient to isolate the divergent part of the distribution,
$$
 {r^ar^b\over (r^2+\ve^2)^5}.
$$
To this order we must apply it to a test function $\vp(\vec r)$,
\be\label{int2}
\left({r^ar^b\over (r^2+\ve^2)^5}\right)(\vp)=\int {r^ar^b \,\vp({\vec r}) \over (r^2+\ve^2)^5}\, d^5r=
{1\over \ve^3} \int {r^a r^b \,\vp(\ve\,{\vec r}) \over (r^2+1)^5}\, d^5r.
\ee
To determine the divergent part for $\ve\ra 0$ we expand $\vp(\ve\,{\vec r})$ in a power series,
\be\label{power}
 \vp(\ve\,{\vec r})=\vp(0)+\ve \,r^a\pa_a\vp(0)+{1\over 2}\,\ve^2\,r^ar^b\,\pa_a\pa_b\vp(0)+o\left(\ve^3\right).
\ee
Due to symmetric integration only the monomials with an even number of $r^a$ survive, and therefore -- due to the presence of the pole  $1/\ve^3$ in \eref{int2} -- only the first and the third terms of the expansion \eref{power} give rise to divergent contributions. The divergent part of \eref{int2} amounts therefore to,
\ba
\left({r^ar^b\over (r^2+\ve^2)^5}\right)_{\rm DIV}(\vp)&=&
{1\over \ve^3} \int {r^a r^b\over (r^2+1)^5}\left(\vp(0)+{1\over 2}\,\ve^2\,r^ar^b\,\pa_a\pa_b\vp(0)\right)\, d^5r\nn\\
&=&{\pi^3\over 4\,\G(5)}\left({1\over \ve^3}\,\dl^{ab}\vp(0)+{1\over \ve}\left(\pa^a\pa^b+{1\over 2}\,\dl^{ab}\,\nabla^2\right)\vp(0)\right),\nn
\ea
where $\nabla^2=\pa_a\pa_a$.
Turning to abstract notation this means that the divergent part of $\displaystyle {r^ar^b\over (r^2+\ve^2)^5}$ in the sense of distributions is,
$$
\left({r^ar^b\over (r^2+\ve^2)^5}\right)_{\rm DIV}={\pi^3\over 4\,\G(5)}\left({1\over \ve^3}\,\dl^{ab}+{1\over \ve}\left(\pa^a\pa^b+{1\over 2}\,\dl^{ab}\,\nabla^2\right)\right)\dl^5(\vec r).
$$
This allows immediately to write down the divergent parts of
\eref{t1}--\eref{t3}, $\wh\Th^{\m\n}_\ve\equiv\left(\Th^{\m\n}_\ve\right)_{\rm DIV}$,
\ba
\wh\Th_\ve^{00}&=& \left({e\over \Omega_5}\right)^2
{\pi^3\over 4\,\G(5)}\left({5\over2}\,{1\over \ve^3}+{7\over 4}\,{1\over \ve}\,\nabla^2\right)\dl^5(\vec r),\label{t1div}
\\
\wh\Th_\ve^{0a}&=& 0,\\
\wh\Th_\ve^{ab}&=& \left({e\over \Omega_5}\right)^2
{\pi^3\over 4\,\G(5)}\left({3\over 2}\,{1\over \ve^3}\,\dl^{ab}+{1\over \ve}\left(-\pa^a\pa^b +{5\over 4}\,\dl^{ab}\nabla^2\right)\right)\dl^5(\vec r).
\ea
In generic flat coordinates these expressions combine to a covariant form -- as they must:
\be\label{cou}
\wh\Th_\ve^{\m\n}=\left({e\over \Omega_5}\right)^2
{\pi^3\over 4\,\G(5)}
\left[{1\over \ve^3}\left(-{3\over 2}\,\eta^{\m\n} +4P^{\m\n}\right)+
{1\over \ve}\left(\left({5\over 4}\,\eta^{\m\n}-3P^{\m\n} \right)\Box-\pa^\m\pa^\n\right)\right]\dl^5,
\ee
where $P^{\m\n}=U^\m U^\n$,  $U^\m=dx^\m/d\s$ and $\Box =\pa_\m \pa^\m$.  Recall also that
for a flat brane  $\Box\, \dl^5=-\nabla^2\,\dl^5$.

\eref{cou} is the counterterm that must be subtracted from \eref{theg}, such that the distributional limit,
\be\label{tmn1}
\wt T^{\m\n}={\rm Lim}\,_{\ve\ra 0}\left(\Th^{\m\n}_\ve - \wh\Th_\ve^{\m\n}\right),
\ee
exists. Notice that \eref{cou} has exactly the structure \eref{re}, because in the present case $\sqrt{g}=1$, $R^{\m\n}_\ve$ is independent of $\s$ and can be brought out of the integral, and the $\dl$--function \eref{dn} can be written as in \eref{dns}.

\subsection{Conservation}
\label{cot}

Formula \eref{tmn1} represents a well--defined distribution, and hence
the divergence $\pa_\m \wt T^{\m\n}$ is also well--defined. According to the strategy of paragraph \ref{dot} we must now evaluate this divergence explicitly and see whether in \eref{tren} we need a finite counterterm $D^{\m\n}$ or not. We show now that in this case $\pa_\m \wt T^{\m\n}$ vanishes. This means that  $D^{\m\n}=0$ and
$
T^{\m\n}=\wt T^{\m\n}.
$

\ul{\it Proof of $\pa_\m \wt T^{\m\n}=0$.}
To compute $\pa_\m \wt T^{\m\n}$ we take advantage from the fact that in the space of distributions the derivative is a {\it continuous} operation. This means that we are allowed to interchange the operations ``${\rm Lim}_{\,\ve\ra 0}$'' and ``$\pa_\m$''. From \eref{tmn1} we get therefore,
\be\label{dtmn1}
\pa_\m \wt T^{\m\n}={\rm Lim}\,_{\ve\ra 0}\left(\pa_\m\Th^{\m\n}_\ve - \pa_\m\wh\Th_\ve^{\m\n}\right).
\ee
From \eref{cou} it is straightforward to compute,
\be\label{dtdiv}
\pa_\m\wh\Th_\ve^{\m\n}=\left({e\over \Omega_5}\right)^2
{\pi^3\over 8\,\G(5)}\left(-{3\over \ve^3}+{1\over2\ve}\,\Box\right)\pa^\n\dl^5.
\ee
The divergence of $\Th_\ve^{\m\n}$ has been obtained in \eref{divteg}, but it can be derived also from the static--coordinate--version \eref{t1}--\eref{t3},
$$
\pa_\m \Th^{\m\n}_\ve=-\left({e\over \Omega_5}\right)^2 \pa^\n \left[{\ve^2\over 2(r^2+\ve^2)^5}\right].
$$
\eref{dtmn1} can then be rewritten as,
\ba
\pa_\m \wt T^{\m\n}&=& \left({e\over \Omega_5}\right)^2
\,{\rm Lim}\,_{\ve\ra 0}\,\pa^\n {\cal F}_\ve,\label{limfe}\\
{\cal F}_\ve&\equiv& -{\ve^2\over 2(r^2+\ve^2)^5}
+{\pi^3\over 8\,\G(5)}\left({3\over \ve^3}-{1\over2\ve}\,\Box\right)\dl^5.\label{fe5}
\ea
Applying ${\cal F}_\ve$ to a test function and proceeding as in \eref{power}, it is straightforward to show that $\lim_{\,\ve\ra0}{\cal F}_\ve(\vp)=0$, $\forall\,\vp$, see subsection \ref{tlo} of appendix B. This means that,
$$
{\rm Lim}\,_{\ve\ra 0}\,{\cal F}_\ve=0.
$$
Since in the space of distributions the derivative is a continuous operation, from \eref{limfe} we conclude then that $
 \pa_\m \wt T^{\m\n}=0$, and no finite counterterm is required.

\subsection{Finite energy integrals}\label{fdm}

Once $T^{\m\n}$ is a distribution, it entails automatically finite energy--momentum integrals $P^\m_V$, see paragraph \ref{fem}. We illustrate this property computing the energy $E_V= P^0_V $ contained in a sphere of radius $R$, surrounding a static particle placed in the origin. For this purpose we must insert \eref{t1} and \eref{t1div} in \eref{pmom}, setting in the latter $\m=0$ and $D=6$,
\ba
E_V&=&
\lim_{\ve\ra 0}\int_V  \left(\Th^{00}_\ve-\wh\Th_\ve^{00}\right)d^5r\nn\\
&=&\left({e\over \Omega_5}\right)^2 \lim_{\ve\ra 0}\int_{r<R}\left[{r^2\over 2\, (r^2+\ve^2)^5}-{\pi^3\over 4\,\G(5)}\left({5\over2}\,{1\over \ve^3}+{7\over 4}\,{1\over \ve}\,\nabla^2\right)\dl^5(\vec r)\right]d^5r.\label{ep6}
\ea
The derivative term, i.e. the laplacian $\nabla^2$, does not contribute to the integral since the test function $\chi_V(\vec x)$ is constant in the neighborhood of the origin. Noting the integral,
$$
\int_0^\infty {r^2\,d^5r\over (r^2+\ve^2)^5}={5\pi^3\over 4\,\G(5)}{1\over \ve^3},
$$
the energy in the sphere becomes then,
\ba
E_V&=&{1\over 2}\left({e\over \Omega_5}\right)^2 \lim_{\ve\ra 0}\left(\int_{r<R}{r^2\,d^5r \over (r^2+\ve^2)^5} -{5\pi^3\over 4\,\G(5)}{1\over \ve^3}\right)=-{1\over 2}\left({e\over \Omega_5}\right)^2 \lim_{\ve\ra 0}\int_{r>R}{r^2\,d^5r \over (r^2+\ve^2)^5}\nn\\
&=&-{1\over 2}\left({e\over \Omega_5}\right)^2\int_{r>R}{d^5r \over r^8}=-{e^2\over 16\pi^2R^3},\nn
\ea
and is thus finite.
Notice that, while the bare energy density $\Th^{00}$ of the field is positive definite -- see \eref{t1} for $\ve=0$ -- the renormalized energy $E_V$ is finite, but negative; this is due to the fact that the counterterm $-\wh\Th_\ve^{00}$ gives a negative and divergent contribution.

From this calculation we see also that the derivative--terms in \eref{cou} are unessential for making the momentum integrals finite (at least in the case of a particle); nevertheless, they are crucial for, 1)  turning $T^{\m\n}$ into a distribution and, 2) for making $T^{\m\n}$ conserved.

\section{String in $D=4$}\label{s4}

In this section we construct the renormalized energy--momentum tensor for the field of a flat string in $D=4$, for which the codimension is even, $n=2$. With respect to the previous case there are two essential new features appearing: 1) the occurrence of a  finite counterterm $D^{\m\n}$ and, 2) the appearance of a dimensionful parameter $l$, that leads to a finite redefinition of the string tension.

A flat string in four dimensions is parametrized  by $x^\m(\s)=U^\m_i\s^i$, where $i=0,1$. In static coordinates $x^\m=(y^i,r^a)$, $i=0,1$, $r=1,2$, disposing the string along the $x^3$--axis we have,
\be\label{sstat}
y^0=x^0,\quad y^1=x^3, \quad   r^1=x^1,\quad r^2=x^2,
\ee
and the parametrization becomes $r^a(\s)=0$, $y^i(\s)=\s^i$.

\subsection{Construction of the divergent counterterm}

We begin again by writing the components of the regularized energy--momentum tensor \eref{theg} in static coordinates ($\Omega_2 =2\pi$),
\ba
\Th_\ve^{ij}&=& {e^2\over 8\pi^2}\,{r^2\,\eta^{ij}\over (r^2+\ve^2)^2},\label{t1s}\\
\Th_\ve^{ia}&=& 0,\\
\Th_\ve^{ab}&=& {e^2\over 8\pi^2}\,{r^2\,\dl^{ab}-2\,r^ar^b\over (r^2+\ve^2)^2}.\label{t3s}
\ea
This time we need thus the divergent part for $\ve\ra 0$ of,
\be\label{div2}
\left({r^ar^b\over (r^2+\ve^2)^2}\right)(\vp)=\int {r^ar^b \,\vp({\vec r}) \over (r^2+\ve^2)^2}\, d^2r=-\pi\ln\ve \,\dl^{ab}\,\vp(0) +constant+o(\ve),
\ee
giving,
$$
\left({r^a r^b\over (r^2+\ve^2)^2}\right)_{\rm DIV}=-\pi\ln\ve \,\dl^{ab}\dl^2(\vec r).
$$
The divergent parts of \eref{t1s}--\eref{t3s} become then,
\ba
\wh\Th_\ve^{ij}&=& -{e^2\over 4\pi}\,\ln\left({\ve\over l}\right)\eta^{ij}\, \dl^2(\vec r),\label{111}\\
\wh\Th_\ve^{ia}&=&0,\\
\wh\Th_\ve^{ab}&=&0,
\ea
or, in generic flat coordinates,
\be\label{th1s}
\wh\Th_\ve^{\m\n}=-{e^2\over 4\pi}\,\ln\left({\ve\over l}\right) P^{\m\n}\,\dl^2.
\ee
The projector $P^{\m\n}$ is defined in \eref{proj}, but in the present case it is constant. In \eref{111}
we have introduced a parameter $l$ with the dimension of length, that is required for dimensional reasons since also $\ve$ has the dimension of length. Notice, however, that the term proportional to $\ln l$ in \eref{th1s} amounts merely to a redefinition of the string tension -- an arbitrariness anticipated in paragraph \ref{uni}. We will come back to this point in subsection \ref{consg}.

\subsection{Construction of a conserved energy--momentum tensor}\label{coa}

As before we have now the well--defined distributional limit,
\be\label{tmns}
\wt T^{\m\n}={\rm Lim}\,_{\ve\ra 0}\left(\Th^{\m\n}_\ve - \wh\Th_\ve^{\m\n}\right),
\ee
and we must evaluate $\pa_\m\wt T^{\m\n}$ to check whether we need a finite counterterm $D^{\m\n}$ or not. To compute $\pa_\m\wt T^{\m\n}$ we take again $\pa_\m \Th_\ve^{\m\n}$ from \eref{divteg},
$$
\pa_\m \Th^{\m\n}_\ve=- {e^2\over 8\pi^2} \,\pa^\n \left({\ve^2\over (r^2+\ve^2)^2}\right).
$$
This time, however, we have the finite distributional limit,
$$
{\rm Lim}_{\,\ve\ra 0}\,\,{\ve^2\over (r^2+\ve^2)^2}=\pi\,\dl^2({\vec r}),
$$
and therefore,
$$
{\rm Lim}_{\,\ve\ra 0}\,\pa_\m\, \Th^{\m\n}_{\ve}=-{e^2\over 8\pi}\,\pa^\n \dl^2.
$$
On the other hand, from \eref{th1s} we have trivially,
$$
\pa_\m\wh\Th_\ve^{\m\n}=0,
$$
since $P^{\m\n}\pa_\m$ is the derivative w.r.t. the tangential coordinates $y^i$.
We obtain therefore,
\be\label{pat}
\pa_\m \wt T^{\m\n}=-  {e^2\over 8\pi}\,\pa^\n \dl^2.
\ee
Comparing with \eref{dwtmn} we get then the {\it operatorial} self--force,
$$
\wt f^\m ={e^2\over 8\pi}\,\pa^\m,
$$
that must be eliminated adding a finite counterterm of the form \eref{dmn}. The  counterterm that does the job is simply,
\be\label{dmn0}
D^{\m\n}={e^2\over 8\pi}\,\eta^{\m\n}\dl^2.
\ee
In fact, since $\pa_\m(\wt T^{\m\n}+D^{\m\n})=0$, the self--force $f^\m$ is algebraic and vanishing -- as must happen for a flat brane.

In conclusion, from \eref{th1s} and \eref{tmns} we obtain the renormalized energy--momentum tensor,
\be
T^{\m\n}=\wt T^{\m\n}+D^{\m\n}=
{\rm Lim}\,_{\ve\ra 0}\left[\Th^{\m\n}_\ve +{e^2\over 4\pi}\left(\ln\left({\ve\over l}\right)P^{\m\n}+{1\over 2}\,\eta^{\m\n}\right)\dl^2\right],\label{tmnsf}
\ee
and,
$$
\pa_\m T^{\m\n}=0.
$$

\subsection{Finite energy integrals}\label{fei}

Since $T^{\m\n}$ in \eref{tmnsf} is a distribution, the four--momentum integrals over finite volumes -- encircling the string or not -- are finite. We illustrate this property again in the case of a static string disposed along the $z$--axis, see \eref{sstat}.

We compute the energy in a cylinder of radius $R$ and of length $L$, concentric with the $z$--axis. The energy in this volume $V$ is obtained inserting \eref{t1s}, \eref{111} and \eref{dmn0} in \eref{pmom}, with $D=4$ and $\m=0$. Equivalently one might integrate the $00$--component of \eref{tmnsf}:
\ba
E_V&=& \lim_{\ve\ra 0} \int_0^L dz\int_{r<R} d^2r\left(\Th_\ve^{00} -\wh\Th_\ve^{00}+ D^{00}\right)  \nn\\
&=&{e^2\over 4\pi^2}\,
\lim_{\ve\ra 0}\int_0^L dz\int_{r<R} d^2r \left[{r^2\,\eta^{00}\over 2 (r^2+\ve^2)^2}
+\pi\left(\ln\left({\ve\over l}\right)\,P^{00}+{1\over 2}\,\eta^{00}\right)\dl^2(\vec r)\right]\nn\\
&=&{e^2L\over 4\pi}\,\lim_{\ve\ra 0} \left(\int_0^R  {r^3\,dr \over (r^2+\ve^2)^2}
+\ln\left({\ve\over l}\right)+{1\over 2}\right)
={e^2L\over 8\pi}\,\lim_{\ve\ra 0}\left(\ln\left({R^2+\ve^2\over l^2}\right)+{\ve^2\over R^2+\ve^2}\right)
\nn\\
&=& {e^2L\over 4\pi}\ln\left({R\over l}\right).
\ea
The energy contained in the volume $V$ is thus finite, but depends on the length scale $l$. As we will see in the next section, this feature occurs for any brane with codimension $n=2$.

\section{The general case}\label{general}

We apply now our approach to a generic flat $(p-1)$--brane in $D$ dimensions. Since the technical details are a bit complicated we relegate them to appendix B, and give here mainly the results. A characteristic feature of the general formulae is that the analytic structure of the  counterterms depends only on the codimension $n$, and not separately on $D$ and $p$. Of course, the results of this section reproduce as particular cases the ones of sections \ref{p6} and \ref{s4}.

\subsection{The divergent counterterm}\label{tdc}

For a generic flat brane the counterterms are made out of a finite number of terms involving derivatives of the $\dl$--function on the brane \eref{dn}, that in flat coordinates reads,
$$
\dl^n=\int\,\dl^D(x-U\cdot\s)\,d^p\s.
$$
The divergent part of the tensor $\Th_\ve^{\m\n}$ in \eref{theg} is then given by (for the derivation see subsection \ref{dotc} of appendix B),
\be\label{divg}
\wh\Th_\ve^{\m\n}=-\left({e\over \Omega_n}\right)^2 \,\sum_{j=0}^{n-2}{}'\, A_n^j\,
\left\{\left[Q^{\m\n}-(n+j)\left(P^{\m\n}-{1\over2}\,\eta^{\m\n}\right)\right]
\Box^{j/2}
-j\,\pa^\m\pa^\n\Box^{j/2-1}
\right\}\dl^n,
\ee
where the ``prime'' indicates that the sum is restricted to {\it even} values of $j$. The coefficients $A_n^j$ are singular for $\ve\ra 0$, being given by,
\be
A_n^j=\cases
{
\displaystyle {(-)^{j/2}\,\pi^{n/2}\,\G\left({n-j\over 2}-1\right)\over 2^{j+1}\,\G(n)\,\G\left({j\over 2}+1\right)}\cdot{1\over \ve^{n-j-2}}, & for $j< n-2$, \cr
&\cr \label{boh}
\displaystyle {(-)^{n/2}\,\pi^{n/2}\over 2^{n-2}\,\G(n)\G\left({n\over2}\right)  }\cdot\ln \left({\ve\over l}\right), & for $j=n-2$.\cr
}
\ee
The coefficient $A_n^{n-2}$ depends on a parameter $l$ with the dimension of length, that has been introduced again for dimensional reasons. Actually,  for $n$ odd  $\wh\Th_\ve^{\m\n}$ is independent of $l$, because the sum in \eref{divg} is only over even $j$. We write explicit expressions for  $n=2,3,4,$ (recall that $Q^{\m\n}=P^{\m\n}-\eta^{\m\n}$),
\be
\wh\Th_\ve^{\m\n}=\cases
{
\displaystyle -{e^2\over 4\pi}\ln\left({\ve\over l}\right)\, P^{\m\n}\,\dl^2, & for $n=2$,
\cr
\cr \label{n23}
\displaystyle {e^2\over 32\ve}\left[P^{\m\n}-{1\over 4}\,\eta^{\m\n}\right]\dl^3, &  for $n=3$,
\cr
\cr
\displaystyle {e^2\over 48\pi^2}\left[{1\over\ve^2}\Big(3P^{\m\n}-\eta^{\m\n}\Big)
+\ln\left({\ve\over l}\right)\left({5\over2}\,P^{\m\n}\,\Box-\eta^{\m\n}\,\Box+\pa^\m\pa^\n\right)\right]\dl^4, & for $n=4$.
\cr
}
\ee
For $n=2$ one recovers the result \eref{th1s} for a string in $D=4$. For $n=3$ one discovers that in $D=4$ the divergent counterterm for a particle in uniform motion coincides with the exact one \eref{p4div} for a particle in {\it arbitrary} motion. We will further comment on this point in section \ref{dyn}.
For $n\ge 4$ $\wh\Th_\ve^{\m\n}$ contains also terms with derivatives of the $\dl$--function; see also \eref{cou} for $n=5$.

By construction the limit,
\be\label{tmnt}
\wt T^{\m\n}={\rm Lim}\,_{\ve\ra 0}\left(\Th^{\m\n}_\ve - \wh\Th_\ve^{\m\n}\right),
\ee
exists, and defines a distribution.

\ul{\it Leading divergences and tension renormalization.}
The term with the leading divergence in \eref{divg}, i.e. the one that is most singular as $\ve\ra 0$, corresponds to $j=0$ and carries no derivatives of the $\dl$--function (while the terms corresponding to subleading divergences contain derivatives). For $n>2$ the leading divergence is a pole $1/\ve^{n-2}$, and the corresponding term is,
\be
\left.\wh\Th_\ve^{\m\n}\right|_{lead}=
 -\left({e\over \Omega_n}\right)^2{\pi^{n/2}\G\left({n\over 2}-1\right)\over2\G(n)\,\ve^{n-2}}\left[(1-n)\,P^{\m\n}+\left({n\over 2}-1\right)\eta^{\m\n}\right]\dl^n.
\label{lead}
\ee
For $n=2$ (and $n=3$) $\wh\Th_\ve^{\m\n}$ contains only leading divergences, and no subleading ones, so that,
$$
\left.\wh\Th_\ve^{\m\n}\right|_{lead}=\wh\Th_\ve^{\m\n}.
$$
In general all terms of $\wh\Th_\ve^{\m\n}$ must be subtracted ``by hand'' from $\Th_\ve^{\m\n}$, to obtain a well--defined energy--momentum tensor, in the sense that they can not be eliminated ``renormalizing'' some fundamental constants of the theory. For $n>2$ even the leading divergence \eref{lead} can not be eliminated renormalizing some coupling constant. Actually, the term proportional to $P^{\m\n}$ could be eliminated renormalizing the tension of the brane, see \eref{tmn}, while the term proportional to $\eta^{\m\n}$ could not be eliminated this way.

On the contrary, for $n=2$ the whole $\wh\Th_\ve^{\m\n}$
{\it could} be eliminated through a  renormalization of the brane tension, since in this case $\wh\Th_\ve^{\m\n}$  has the same structure as the energy--momentum tensor of the brane \eref{tmn}. We stress, however, that our general philosophy does not rely on such a renormalization, since we subtract all divergences ``by hand''.

\subsection{Finite counterterms and energy--momentum conservation}\label{consg}

To evaluate $\pa_\m\wt T^{\m\n}$ we need the divergence of $\Th^{\m\n}_\ve$ in \eref{divteg},
\be\label{divteg0}
\pa_\m\Th^{\m\n}_\ve=-\left({e\over \Omega_n}\right)^2 \pa^\n {\ve^2 \over 2\left(Q_{\a\bt}x^\a x^\bt+\ve^2\right)^n},
\ee
and the divergence of \eref{divg},
\be
\label{ddivg}
\pa_\m\wh\Th_\ve^{\m\n}=-{1\over2} \left({e\over \Omega_n}\right)^2 \,\sum_{j=0}^{n-2}{}' \, (n-j-2)\,A_n^j\,\pa^\n \,\Box^{j/2}\,\dl^n.
\ee
From \eref{tmnt} we obtain then,
\ba
\pa_\m \wt T^{\m\n}&=& \left({e\over \Omega_n}\right)^2
\,{\rm Lim}\,_{\ve\ra 0}\,\pa^\n {\cal F}_\ve,\label{limfeg}\\
{\cal F}_\ve&\equiv& -{\ve^2\over 2(Q_{\a\bt}x^\a x^\bt+\ve^2)^n}
+{1\over2}\,\sum_{j=0}^{n-2}{}'\,(n-j-2)\,A_n^j\,\Box^{j/2}\,\dl^n.\label{feg}
\ea
The distributional limit ${\rm Lim}\,_{\ve\ra 0}\,{\cal F}_\ve$ is evaluated in subsection \ref{tlo} of appendix B, and for {\it even} $n$ it is non--vanishing, see \eref{b12}. It turns then out that,
\be\label{dtilde}
\pa_\m \wt T^{\m\n}=\cases
{
\displaystyle \left({e\over \Omega_n}\right)^2\pa^\n\left(
{(-)^{n/2}\,\pi^{n/2}\over 2^{n-1}\,\G(n)\G\left({n\over2}\right)}
\,\Box^{n/2-1}\, \dl^n\right), &for $n$ even,
\cr
&\cr
0, &for $n$ odd.
\cr
}
\ee
Comparing with \eref{dwtmn} we conclude that for $n$ even the self--force $\wt f^\m$ is operatorial, but by inspection we see that it can be  eliminated through the finite counterterm,
\be
 D^{\m\n}=\cases
{
\displaystyle -\left({e\over \Omega_n}\right)^2
{(-)^{n/2}\,\pi^{n/2}\over 2^{n-1}\,\G(n)\G\left({n\over2}\right) }
\,\eta^{\m\n}\, \Box^{n/2-1}\, \dl^n, &for $n$ even,
\cr
&\cr\label{dwt}
 0, &for $n$ odd.
\cr
}
\ee
The conserved energy--momentum tensor is then,
$$
T^{\m\n}=\wt T^{\m\n} +D^{\m\n},\quad\quad
\pa_\m T^{\m\n}=0, \quad\quad f^\m=0.
$$
A non--vanishing finite counterterm is thus required for every brane with even codimension.

\ul{\it Uniqueness.} As shown in paragraph \ref{uni}, the energy--momentum tensor is unique, modulo a redefinition of the brane tension, and modulo the shifts,
\be\label{ambi0}
T'^{\m\n} = T^{\m\n} +\pa_\rho V^{\rho\m\n},\quad\quad \mbox{with}\quad
V^{\rho\m\n}=-V^{\m\rho\n}.
\ee
On the contrary, our $T^{\m\n}$ derived above, in the case of even $n$ does not seem to be unique, since it depends on an arbitrary parameter $l$. More precisely,
the contribution of $T^{\m\n}$ depending on $l$ comes from the term with $j=n-2$ in \eref{divg}, and amounts to,
$$
T^{\m\n}_l=\left({e\over \Omega_n}\right)^2
{(-)^{n/2}\,\pi^{n/2}\,\ln l\over 2^{n-2}\,\G(n)\G\left({n\over2}\right)}\,
\Big[
(2n-3)\,P^{\m\n}\,\Box+(2-n)\,(\eta^{\m\n}\Box-
\pa^\m\pa^\n)\Big]\Box^{n/2-2}\dl^n.
$$
For consistency
$T^{\m\n}_l$ must then be absorbable through a redefinition of the brane tension, or through a shift of the kind \eref{ambi0}. This is indeed the case. In fact, if $n=2$ we have $T^{\m\n}_l\propto P^{\m\n}\,\dl^2$, which corresponds to a shift of the brane tension. On the other hand, if $n\ge 4$ we have
$
T^{\m\n}_l=\pa_\rho V^{\rho\m\n}_l,
$
where,
\ba
V^{\rho\m\n}_l&=&\left({e\over \Omega_n}\right)^2
{(-)^{n/2}\,\pi^{n/2}\,\ln l\over 2^{n-2}\,\G(n)\G\left({n\over2}\right)}\,
\Big[
(2n-3)\,(\pa^\rho P^{\m\n}-\pa^\m P^{\rho\n})\nn\\
&&\phantom{ rrrrrrdddddddddddddd}+(2-n)\,(\eta^{\m\n}\pa^\rho-\eta^{\rho\n}\pa^\m)\Big]\Box^{n/2-2}\dl^n.\nn
\ea

In conclusion, if $n$ is odd the energy--momentum tensor is given by $T^{\m\n}=\wt T^{\m\n}$, and it is uniquely determined. If $n$ is even it is given by $T^{\m\n}=\wt T^{\m\n}+ D^{\m\n}$, and it depends on the dimensionful parameter $l$. For $n=2$ this parameter corresponds to a shift of the brane tension (see \cite{DQ,CHH} for the appearance of a similar parameter in the self--force of a string in $D=4$),
while for $n\ge4$ this dependence is spurious, in the sense of \eref{ambi0}.

\subsection{Finite energy integrals}

To illustrate the finiteness of the momentum integrals we take a static $(p-1)$--brane disposed along the coordinates $y^1,\cdots,y^{p-1}$, see \eref{stat2}, \eref{stat3}, and compute the energy $E_V$ in a tubular volume $V$ containing a hypercubic portion of the brane of volume $L^{p-1}$, delimited by $r=\sqrt{r^ar^a} <R$,  $0<y^i<L$ $(i=1,\cdots,p-1)$.

The energy in this volume is given by \eref{pmom} with $\m=0$. We need thus in particular the regularized energy density $\Th_\ve^{00}$, that can be read off from \eref{theg},
\be\label{th00}
\Th_\ve^{00}=\left({e\over \Omega_n}\right)^2 {r^2\over 2(r^2+\ve^2)^n}.
\ee

\ul{\it $n>2$.} We consider first a brane with codimension $n>2$. In this case the finite counterterm  \eref{dwt} does not contribute to the integral \eref{pmom}, because it is a derivative term supported on the brane, and on the brane the test function $\chi_V(\vec x)$ is constant. For the same reason the divergent counterterm \eref{divg} contributes only with its derivative--free leading term \eref{lead},
$$
\wh\Th_\ve^{00}\quad\ra\quad
\left.\wh\Th_\ve^{00}\right|_{lead}=
 \left({e\over \Omega_n}\right)^2{n\,\pi^{n/2}\G\left({n\over 2}-1\right)\over4\G(n)\,\ve^{n-2}}\,\dl^n(\vec r).
$$
From \eref{pmom} we get then,
\ba
E_V&=& \lim_{\ve\ra0}\,\,\prod_{i=1}^{p-1}\int_0^Ldy^i\,\int_{r<R} d^nr\left(\Th_\ve^{00} -\left.\wh\Th_\ve^{00}\right|_{lead}\right)  \nn\\
&=&
L^{p-1}\left({e\over \Omega_n}\right)^2
\lim_{\ve\ra 0}\left[\int_{r<R} {r^2d^nr\over 2(r^2+\ve^2)^n}-
{n\,\pi^{n/2}\G\left({n\over 2}-1\right)\over4\G(n)\,\ve^{n-2}}\right]\nn\\
&=&
L^{p-1}\left({e\over \Omega_n}\right)^2
\lim_{\ve\ra 0}\left[-\int_{r>R}{r^2d^nr\over 2(r^2+\ve^2)^n}
+\int_0^\infty {r^2d^nr\over 2(r^2+\ve^2)^n}
-{n\,\pi^{n/2}\G\left({n\over 2}-1\right)\over4\G(n)\,\ve^{n-2}}\right].\nn
\ea
The last two terms of the square brackets compensate exactly -- even before taking the limit $\ve\ra 0$ -- and in the first term the limit can now be taken trivially. The result is therefore,
\be\label{evg}
E_V=-L^{p-1}\left({e\over \Omega_n}\right)^2
\int_{r>R}{r^2d^nr\over 2\,r^{2n}}=-{e^2\over 2\,(n-2)\,\Omega_n}\cdot{L^{p-1}\over R^{n-2}}.
\ee
Again the renormalized energy $E_V$ is negative. Nevertheless, $E_V$ is an increasing function of the radius $R$.

\ul{\it $n=2$.} For $n=2$ all three terms of \eref{pmom} contribute to the integral, since in this case the finite counterterm \eref{dwt} does not contain derivatives.
Inserting \eref{n23}, \eref{dwt} and  \eref{th00} in \eref{pmom}, one gets (this is the same calculation as in subsection \ref{fei}),
\ba
E_V&=& \lim_{\ve\ra0}\,\,\prod_{i=1}^{p-1}\int_0^Ldy^i\, \int_{r<R} d^2r\left(\Th_\ve^{00} -\wh\Th_\ve^{00}+ D^{00}\right) \nn\\
&=&L^{p-1} \left({e\over 2\pi}\right)^2
\lim_{\ve\ra 0}\int_{r<R} d^2r \left[{r^2\over 2\, (r^2+\ve^2)^2}+\pi\ln
\left({\ve\over l}\right)
\,\dl^2(\vec r)+{\pi\over2}\,\dl^2(\vec r)\right]
\nn\\
&=&
{e^2L^{p-1}\over4\pi}\ln\left({R\over l}\right).
\ea

\section{Branes in arbitrary motion: a preliminary analysis} \label{dyn}

For a brane in arbitrary motion the main new aspect of our approach for the construction of a consistent energy--momentum tensor, is that it provides a new paradigm for the derivation of the self--force.
The first step of the derivation is the identification of the counterterms that make, in particular, the limit
$\wt T^{\m\n}={\rm Lim}\,_{\ve\ra 0}\left(\Th^{\m\n}_\ve-\wh\Th_\ve^{\m\n}\right)$ well--defined.
While we hope to furnish explicit derivations of the self--force in the future, in this final section we perform a preliminary analysis of the new counterterms involved, if the brane is no longer flat but in arbitrary motion.

The conclusion of this section is that new {\it subleading} divergent counterterms, and  new finite counterterms, might indeed show up if $n\ge 4$. This does not contradict our uniqueness theorem of paragraph \ref{uni}, since these new counterterms vanish in the case of a flat brane.

\subsection{Generalized curvatures}

From a geometrical point of view the difference between a flat brane and a brane in arbitrary motion is that for the latter the generalized velocities are no longer (covariantly) constant,
$$
U_{ij}^\m\equiv D_iU_j^\m\neq0, \quad\quad U_{ij}^\m=U_{ji}^\m.
$$
This means that the geometry on the brane is now enriched by the presence of the ``generalized curvatures'',
\be\label{curv}
U_{(p)}\equiv U_{i_1\cdots \,i_p\,i_{p+1}}^\m=D_{i_1}\cdots D_{i_p}U^\m_{i_{p+1}},
\ee
that may give rise to new contributions to the divergent and finite counterterms
of a flat brane, \eref{divg} and \eref{dwt}. These new contributions  modify the tensors $R^{\m\n}_\ve$ and $\D^{\m\n}$ in \eref{re}, \eref{dmn}, by terms that schematically are of the form \footnote{We assume here -- as a preliminary analysis suggests -- that also for a brane in arbitrary motion the tensors $R^{\m\n}_\ve(\s)$ and $\D^{\m\n}(\s)$ are ``local'', in the sense that they depend on the generalized curvatures evaluated at the same point $\s$.},
\be \label{mnk}
I_{mqk}\equiv {1\over \ve^m}\,U_{(p_1)}\cdots U_{(p_N)}\,(\pa_\m)^k,\quad q=p_1+\cdots+p_N\ge0,   \quad m\ge0,\quad k\ge 0.
\ee
Here it is understood that $1/\ve^0$ stands for $c_1\ln \ve+c_2$, and that the contractions of indices occur through the tensors $U^\m_i$ and $\eta^{\m\n}$, or their combinations $P^{\m\n}$ and $g_{ij}$. The integer $q$ is the total number of derivatives on the $U^\m_i$. The new {\it finite} counterterms arise from $m=0$.

The presence or absence of terms like \eref{mnk} is highly constrained by covariance and dimensional reasons. The length--dimension of $I_{mqk}$ is, in fact, $1/L^{m+q+k}$.
Moreover, since we are searching for tensors $R^{\m\n}_\ve$ and $\D^{\m\n}$ with an even number (i.e. two) of indices, and since $U_{(p)}$ involves $p$ derivatives of  $U^\m_i$, we obtain as first constraint that $q+k$ must be even. A second constraint derives from the fact that the terms with $q=0$  -- that do not involve  generalized curvatures -- are of the same form as the counterterms \eref{divg} and \eref{dwt}, whose coefficients are already fixed. This means that terms $I_{mqk}$ with $q=0$ are not allowed. In conclusion, the tensors $I_{mqk}$ that can modify $R^{\m\n}_\ve$ and $\D^{\m\n}$ must obey,
\be\label{cond}
q+k=\mbox{even}, \quad q\ge1.
\ee
In the following we perform a preliminary analysis of the possible new terms, at fixed codimension.

\subsection{Codimension $n=2$}

For $n=2$, in the flat case the structure of the
counterterms \eref{n23}, \eref{dwt} is simple, and the renormalized energy--momentum tensor is given by, see also \eref{tmnsf},
\be
T^{\m\n}=
{\rm Lim}\,_{\ve\ra 0}\left[\Th^{\m\n}_\ve +{e^2\over 4\pi}\,I^{\m\n}_2
\dl^2\right],\label{n2s}
\ee
where $I^{\m\n}_2$ is the constant tensor,
\be\label{i2}
 I^{\m\n}_2\equiv\ln\left({\ve\over l}\right)P^{\m\n}+{1\over 2}\,\eta^{\m\n}.
\ee
For a brane in arbitrary motion there is a unique way to covariantize the expression \eref{n2s},
\be
T^{\m\n}=
{\rm Lim}\,_{\ve\ra 0}\left[\Th^{\m\n}_\ve +{e^2\over 4\pi}
\int\sqrt{g}\,\,I^{\m\n}_2\,
\dl^D(x-x(\s))\,d^p\s\right],\label{n2}
\ee
where formally $I^{\m\n}_2$ is again given by \eref{i2}, but now $P^{\m\n}(\s)=g^{ij}U_i^\m U_j^\n$ is no longer constant, and hence $I^{\m\n}_2\equiv I^{\m\n}_2(\s)$ is a function  of $\s$. Moreover, in \eref{n2} $\Th^{\m\n}_\ve$ is now the general expression given by \eref{rete}, \eref{ae}, \eref{the}. For a brane in arbitrary motion
the integrand $I^{\m\n}_2$ could -- in principle -- be modified by the addition of tensors of the type \eref{mnk}. But since $I^{\m\n}_2$ is dimensionless and the length--dimension of $I_{mqk}$ is $1/L^{m+q+k}$, we must have $m=q=k=0$, in contrast with \eref{cond}. No such tensors can, therefore, appear.
We conclude that also for a brane in arbitrary motion the distributional limit in \eref{n2} exists: formula \eref{n2} identifies thus the renormalized energy--momentum tensor for a brane in arbitrary motion with codimension $n=2$.

The new aspect is that now  $\pa_\m T^{\m\n}$ is different from zero, and the explicit evaluation of this divergence -- in the sense of distributions -- would then allow to derive the self--force of the brane according to \eref{selff}.

\subsection{Codimension $n=3$}

In this case from \eref{n23} and \eref{dwt} we obtain for a flat
brane,
$$
T^{\m\n}=
{\rm Lim}\,_{\ve\ra 0}\left[\Th^{\m\n}_\ve -{e^2\over 32}\,I_3^{\m\n}
\dl^3\right],
$$
where,
\be\label{imn}
I^{\m\n}_3={1\over \ve}\left(P^{\m\n}-{1\over 4}\,\eta^{\m\n}\right).
\ee
For a brane in arbitrary motion we would thus obtain,
\be\label{n3}
T^{\m\n}={\rm Lim}\,_{\ve\ra 0}\left[\Th^{\m\n}_\ve -{e^2\over 32}\int \sqrt{g}\,\,I^{\m\n}_3\,
\dl^D(x-x(\s))\,d^p\s
\right].
\ee
The tensor $I^{\m\n}_3$ has now dimension $1/L$, and a priori we could then add terms $I_{mqk}$ with $m+q+k=1$. But again \eref{cond} admits no solution. This means that also for $n=3$ no $I_{mqk}$ can show up, and hence the limit \eref{n3} must exist also for a brane in arbitrary motion.

The argument just given is, actually, not directly valid if the brane is a particle in $D=4$. In that case the tensor \eref{imn} could, indeed, be modified by two new covariant terms with the correct dimension. Choosing for $\s^0$ the proper time $\s$ one could, in fact, modify $I_3^{\m\n}$ according to,
$$
 I^{\m\n}_3\quad\ra\quad I^{\m\n}_3+\left(c_1\ln\left({\ve\over l}\right)+c_2\right)\left(U^\m {dU^\n\over d\s}+U^\n {dU^\m\over d\s}\right),
$$
where $c_1$ and $c_2$ are dimensionless constants.
The (divergent) term proportional to $c_1$ is indeed present at intermediate steps of the explicit evaluation of $\wh \Th^{\m\n}_\ve$, but eventually it drops out \cite{LM}. On the other hand, the new (finite) term proportional to $c_2$ would lead to an {\it operatorial} self--force, and it is thus forbidden by energy--momentum conservation. We have thus $c_1=c_2=0$. For a particle in $D=4$ the explicit calculation gives indeed \eref{p4}, \eref{p4div}, in agreement with \eref{n3}.

 \subsection{Codimension $n=4$}

This is the first case where new counterterms $I_{mqk}$ might show up. Proceeding as above \eref{n23} and \eref{dwt} give for a flat
brane,
\be\label{n4s}
T^{\m\n}={\rm Lim}\,_{\ve\ra 0}\left[\Th^{\m\n}_\ve -{e^2\over 48 \pi^2}\,I_4^{\m\n}\,\dl^4\right],
\ee
where,
$$
I_4^{\m\n}={1\over\ve^2}\,\left(3P^{\m\n}-\eta^{\m\n}\right)
+\ln\left({\ve\over l}\right)\left({5\over 2}\,P^{\m\n}\,\Box+\pa^\m\pa^\n-\eta^{\m\n}\Box\right)-{1\over4}\,\eta^{\m\n}\,\Box.
$$
For a brane in arbitrary motion \eref{n4s} would thus lead to,
\be\label{n40}
T^{\m\n}_{(0)}={\rm Lim}\,_{\ve\ra 0}\left[\Th^{\m\n}_\ve -{e^2\over 48\pi^2}\int \sqrt{g}\,\, I_4^{\m\n}\,\dl^D(x-x(\s))\,d^p\s
\right].
\ee
Since $I_4^{\m\n}$ has dimension $1/L^2$, we could now have new counterterms $I_{mqk}$ with $m+q+k=2$. In this case there are, indeed, two possible structures compatible with \eref{cond}, corresponding to ($m=0$, $q=2$, $k=0$) and ($m=0$, $q=1$, $k=1$), that could modify $I_4^{\m\n}$ through a finite number of new terms,
\be\label{in4}
\wt I_4^{\m\n}= I_4^{\m\n}+\left(a_1+b_1\ln\left({\ve\over l}\right)\right) g^{ij}g^{mn}U_{ij}^\m U_{mn}^\n
+\left(a_2+b_2\ln\left({\ve\over l}\right)\right) g^{ij}U^{(\m}_{ij}\pa^{\n)}+\mbox{contractions}.
\ee
With ``contractions'' we mean terms with the same structure as the previous ones, but with different contractions of the indices, and $a_i$ and $b_i$ are dimensionless constants.
The energy--momentum tensor of the field would then be given by,
\be\label{n4}
T^{\m\n}={\rm Lim}\,_{\ve\ra 0}\left[\Th^{\m\n}_\ve -{e^2\over 48\pi^2}\int \sqrt{g}\,\, \wt I_4^{\m\n}\,\dl^D(x-x(\s))\,d^p\s
\right].
\ee
The coefficients $ b_i$ must be determined requiring that the limit in \eref{n4} exists, and this requires to evaluate explicitly the divergent part of $ \Th^{\m\n}_\ve$. The coefficients $a_i$, on the other hand, must be determined demanding that the self--force in \eref{selff} becomes algebraic. A priori we found no stringent reason to assume that all, or some, of these coefficients are zero, even if it seems unlikely that a dependence on $l$ (required by dimensional reasons) as the one in \eref{in4} is compatible with an algebraic self--force.

For a particle in $D=5$ there might be an additional term with the correct dimension, with a pole $1/\ve$, given by,
$$
\wt I^{\m\n}_4\quad \ra\quad  \wt I^{\m\n}_4+{c_1\over \ve}\left(U^\m {dU^\n\over ds}+U^\n {dU^\m\over ds}\right),
$$
but we conjecture that it is absent, as for a particle in $D=4$.

In general, for a generic $n>4$ the dimension of $I_n^{\m\n}$ is $1/L^{n-2}$, and one has then a series of possible new counterterms of the type $\eref{mnk}$, involving higher poles up to $1/\ve^{n-4}$. As mentioned above, the effective appearance of these terms can be checked only through an explicit calculation.

\section{Summary and open problems} \label{sum}

We have proposed a general construction for the renormalized energy--momentum tensor for the  field of a brane in $D$ dimensions -- represented by formulae \eref{tren}--\eref{dmn} -- that leads to a unique result.
The construction does not modify the value of the energy--momentum tensor in the complement of the brane, and preserves all symmetries of the system.
Our proposal is conceptually rater simple and -- we sustain --  of {\it fundamental character}, since it  allows for a new definition of the self--force of the brane that, by construction, ensures automatically energy--momentum conservation. Although conceptually simple, the explicit determination of the self--force may be complicated. In this paper we performed a consistency--check of our proposal, giving a constructive proof that it works for a generic flat brane in $D$ dimensions, where the self--force is zero.

It is rather obvious how to extend our approach to branes that are coupled, in addition to the antisymmetric potential, also to a scalar field $\vp$ and to linearized gravity $h_{\m\n}$, with $g_{\m\n}=\eta_{\m\n}+h_{\m\n}$. In fact, also the solutions for these fields involve the Green--function $G$ of the d'Alembertian, and it is thus sufficient to define the regularized solutions $\vp_\ve$ and $h_{\ve\,\m\n}$ through the replacement $G\ra G_\ve$, see \eref{greene}. The approach proceeds then in the same way as for the coupling to the antisymmetric potential in subsection \ref{rem}. In this way one should be able, for example, to prove for the classical Nambu--Goto superstring, coupled to the antisymmetric tensor, the graviton and the dilaton, the non--renormalization theorems that have been proven so far only for what concerns the leading divergences  \cite{DH,CHH,BD1}.

\ul{\it Open problems.} The main open questions regard branes in arbitrary motion, and first of all  the structure of the divergent counterterm $\wh\Th_\ve^{\m\n}$ in \eref{re}. By construction it is supported on the brane, but we need still a proof of the fact that $R^{\m\n}_\ve(\s)$ is ``local'', in the sense that it depends only on the curvatures \eref{curv} at the same point $\s$. The preliminary analysis of the previous section was, indeed, based on this assumption. Nevertheless, even if this property does not hold -- in which case the structure of the divergent counterterms is more complicated -- our approach maintains its validity.

The second problem regards the (derivation of the) self--force.
The divergence of $\wt T^{\m\n}$ in  \eref{ttilde} has necessarily the structure \eref{dwtmn}, i.e.
$\pa_\m \wt T^{\m\n}= -\int\sqrt{g}\,\wt f^\n \,\dl^D(x-x(\s))\,d^p\s$, but a priori it is by no means obvious that there exists a finite counterterm $D^{\m\n}$, i.e. a tensor $\D^{\m\n}$, that renders the self--force $f^\n=\wt f^\n-\D^{\m\n}\,\pa_\m$ in \eref{algf} algebraic. On the other hand, if no such $\D^{\m\n}$ exists, then we must conclude 1) that our approach is incomplete, or 2)
that classical brane theory is incompatible with energy--momentum conservation.

The third problem regards the fact that the self--force, that in our approach is uniquely determined, must satisfy the constraint,
\be \label{uif}
U^\m_i f_\m=0.
\ee
This property is not guaranteed a priori, and this point deserves thus further investigation. Here we can only recall that for particles and dyons in $D=4$ the constraint \eref{uif} is indeed satisfied \cite{L}.

A last question regards the meaning of the dimensionful parameter $l$ in \eref{boh}: on one hand it is required for dimensional reasons, but on the other its presence is ``spurious'', in that it gives rise to physically equivalent energy--momentum tensors.  The correct interpretation of $l$ can probably be given only in the context of branes in arbitrary motion.

The questions just raised can presumably be addressed only through an explicit evaluation
of the counterterms and of the self--forces. An explicit evaluation of the latter is  also needed to compare the self--forces following from our approach, with the ones obtained previously through conventional techniques \cite{MM,G2,CBU,BCM,KS}. A particle in generic (even and odd dimensional) space--times, as well as a string in $D=4$, seem to represent appropriate laboratories to attack these issues. We plan to address them in a future publication.

\vskip0.5truecm

\paragraph{Acknowledgements.}
This work is supported in part by the INFN Special Initiative TV12.

\vskip0.5truecm

\section{Appendix A: the regularized potential}

In this appendix we derive formulae \eref{aen} for the regularized potential $A^{\m_1\cdots\,\m_p}_\ve$. To this order we must
evaluate the integral appearing in \eref{ae2},
\be\label{int0}
I_\ve \equiv \int G_\ve(x-U\cdot\s)\,d^p\s,\quad\quad x-U\cdot\s\equiv x^\m-U^\m_i\s^i,
\ee
with $G_\ve$ given in \eref{greene}.

\subsection{Even dimensions: $D=2N+4$}

In even dimensions the integral \eref{int0} becomes, using \eref{greene},
$$
 I_\ve={1\over 2\pi^{N+1}}\left(-{d\over d\ve^2}\right)^N \int H\left(x^0-U^0_i\,\s^i\right)\,
\dl\left((x-U\cdot\s)^2-\ve^2\right)\,d^p\s.
$$
To evaluate the integral it is convenient to switch to static coordinates $x^\m\ra(y^i,r^a)$, see \eref{stat1}--\eref{stat4},
$$
I_\ve ={1\over 2\pi^{N+1}}\left(-{d\over d\ve^2}\right)^N \int H\left(y^0-\s^0\right)
\dl\left( (y^0-\s^0)^2-|\vec y-\vec \s|^2 -r^2-\ve^2\right)d^p\s,
$$
where $\vec \s=(\s^1,\cdots,\s^{p-1})$, and similarly for $\vec y$, and $r=\sqrt{r^ar^a}$. Integrating over $\s^0$, and shifting $\vec \s\ra \vec \s+\vec y$, one obtains,
\ba
I_\ve&=&{1\over 4\pi^{N+1}}\left(-{d\over d\ve^2}\right)^N \int {d^{p-1}\s\over \sqrt{|\vec \s|^2 +r^2+\ve^2}}\nn\\
&=&{1\over 4\pi^{N+1}}\,{1\over 2}\cdot{3\over2}\cdots{2N-1\over2} \int {d^{p-1}\s\over (|\vec \s|^2 +r^2+\ve^2)^{N+1/2}}.\label{int1}
\ea
For $n=D-p>2$ the integral in \eref{int1} converges, and it is elementary,
\be\label{ge1}
I_\ve =
{\Gamma\left({n\over 2}-1\right)\over 4\pi^{n/2}\left(r^2+\ve^2\right)^{n/2-1}}.
\ee
For $n=2$ an infrared cut--off, say  $\Lambda$, for the integration variable $\vec \s$ in \eref{int1} is required, due to the infinite extension of the brane \footnote{No cut--off is needed, instead, if one computes directly the field $F_\ve^{\m_1\cdots\,\m_{p+1}}$, since in that case one needs the derivative w.r.t. $r^a$ of \eref{int1}.}. Correspondingly we replace in \eref{int1}  $\int d^{p-1}\s\ra \int_{|\vec\s|<\Lambda} d^{p-1}\s$. Sending $\Lambda$ to  $\infty$ one obtains then,
\be\label{ge2}
I_\ve \quad\ra\quad -{1\over 4\pi}\,\ln{r^2+\ve^2\over \Lambda^2}+c +o\left({1 \over\Lambda}\right),
\ee
where $c$ is a constant.
With \eref{ge1}, \eref{ge2}, and taking into account that for generic flat coordinates $r^2= x^\m x^\n Q_{\m\n}$, \eref{ae2} becomes \eref{aen}.

\subsection{Odd dimensions: $D=2N+3$}

In odd dimensions the integral \eref{int0} becomes,
$$
I_\ve={1\over 2\pi^{N+1}}\left(-{d\over d\ve^2}\right)^N \int H\left(x^0-U^0_i\,\s^i\right)\,
{H(\left(x-U\cdot\s)^2-\ve^2\right)\over
\sqrt{(x-U\cdot\s)^2-\ve^2}}\,d^p\s,
$$
or, in static coordinates, and shifting $\s^i\ra \s^i+y^i$,
\be\label{gedis}
I_\ve ={1\over 2\pi^{N+1}}\left(-{d\over d\ve^2}\right)^N \int H\left(-\s^0\right)\,
{H\left((\s^0)^2-|\vec \s|^2 -r^2-\ve^2\right)
\over
\sqrt{(\s^0)^2-|\vec \s|^2 -r^2-\ve^2}}\,d^p\s.
\ee
To parallel the computation for even $D$ we cut--off the integral over $\s^0$, say, requiring $\s^0>-L$, perform then  one derivative w.r.t. $\ve^2$, send then $L\ra \infty$, perform then the remaining $N-1$ derivatives w.r.t. $\ve^2$, and perform eventually the integration over $\vec\s$. Restricting in \eref{gedis} $\s^0>-L$ and sending then $\s^0\ra-\s^0$, one obtains,
\ba\nn
I_\ve&=&{1\over 2\pi^{N+1}}\left(-{d\over d\ve^2}\right)^N \int d^{p-1}\s \int^L_{\sqrt{|\vec \s|^2 +r^2+\ve^2}}\,
{d\s^0\over
\sqrt{(\s^0)^2-|\vec \s|^2 -r^2-\ve^2}}\\
&=&
{1\over 2\pi^{N+1}}\left(-{d\over d\ve^2}\right)^N \int d^{p-1}\s\,\,{\rm arccosh}\left({L\over \sqrt{|\vec \s|^2 +r^2+\ve^2}}\right).\label{gedis1}
\ea
Since,
\ba
-{d\over d\ve^2}\,{\rm arccosh}\left({L\over \sqrt{|\vec \s|^2 +r^2+\ve^2}}\right)&=&{L\over
2(|\vec \s|^2 +r^2+\ve^2)\sqrt{L^2-|\vec \s|^2 -r^2-\ve^2}}\nn\\
&\ra&{1\over 2 (|\vec \s|^2 +r^2+\ve^2)},\quad\quad\mbox{for}\quad {L\ra\infty},\nn
\ea
\eref{gedis1} becomes,
$$
I_\ve ={1\over 4\pi^{N+1}}\left(-{d\over d\ve^2}\right)^{N-1}
\int {d^{p-1}\s\over |\vec \s|^2 +r^2+\ve^2}.
$$
The remaining derivatives and integrals are now elementary as in \eref{int1} -- for $n=2$ one needs again an infrared cut--off $\Lambda$ -- and the final result is again \eref{ge1}, \eref{ge2}, with $n=D-p$. Also in odd space--time dimensions the regularized potential is therefore given by \eref{aen}.

\section{Appendix B: technical details of section \ref{general}}

\subsection{Determination of the divergent counterterm}\label{dotc}

In this section we derive the general expression \eref{divg} for
the divergent counterterm $\wh\Th_\ve^{\m\n}$. Since  $\wh\Th_\ve^{\m\n}$ is -- by definition --  the divergent part of the tensor $\Th_\ve^{\m\n}$ in \eref{theg}, its determination requires the evaluation of
the divergent part of,
\be\label{b1}
\left({Q^{\m\a}\,Q^{\n\bt}\,x_\a x_\bt \over \left(Q_{\a\bt}\,x^\a x^\bt +\ve^2\right)^n}\right)_{\rm DIV}.
\ee
Since in this expression only the orthogonally projected coordinates $Q^{\m\n}x_\n$  appear, it is sufficient to go to static coordinates ($r^2= Q_{\a\bt}\,x^\a x^\bt$), and to  determine the divergent part of,
\be\label{b2}
\left({r^ar^b \over \left(r^2 +\ve^2\right)^n}\right)_{\rm DIV}.
\ee
To extract from this expression the divergent part one must apply it to a test function $\vp(\vec r)$, and isolate the divergences as $\ve\ra 0$.
We proceed as in \eref{int2}, \eref{power}, sending $\vec r \ra \ve\, \vec r$ and expanding $\vp(\ve\,\vec r)$ in Taylor series. We consider separately the cases of $n$ even and odd.

{\bf 1)  $n$ odd.} In this case we have,
\ba
&&\left({r^ar^b \over \left(r^2 +\ve^2\right)^n}\right)(\vp)=
\int {r^ar^b \vp(\vec r)\over \left(r^2 +\ve^2\right)^n}\,d^nr\label{b3}\\
&=&
{1\over \ve^{n-2}}\,\sum_{j=0}^{n-3} {\ve^j\over j!} \left(\int {r^ar^b\over (r^2+1)^n}\,
r^{a_1}\cdots r^{a_j}\,d^nr\right) \pa_{a_1}\cdots\pa_{a_j}\vp(0) \label{b4}\\
 &&+\mbox {(terms that converge for $\ve \ra 0$}).\nn
\ea
Due to symmetric integration only {\it even} values of $j$ contribute in the sum, and
hence only odd poles $1/\ve^{2m+1}$ appear. In particular no logarithmic divergence $\sim \ln\ve$ shows up.
The integrals over $d^nr$ are elementary. Going to polar coordinates $r^a=r\,n^a$, $d^nr= r^{n-1}dr\,d\Omega_n$, $n^an^a=1$, it is sufficient to remember the standard integrals,
\be\label{in1}
\int d\Omega_n\left(n^{a_1}\cdots n^{a_{2j}}\right)=\Omega_n {1\cdot3\cdots(2j-1)\over n(n+2)\cdots(n+2j-2)}\,\dl^{(a_1a_2}\cdots\dl^{a_{2j-1}\,a_{2j})},
\ee
and,
\be\label{in2}
\int_0^\infty {r^M dr\over (r^2+1)^N}={1\over 2}\,{\G\left({M+1\over 2}\right)\G\left(N-{M+1\over 2}\right)\over \G(N)}.
\ee
Working out the combinatorics, the divergent part of \eref{b4} becomes,
\be\label{b5}
\left({r^ar^b \over \left(r^2 +\ve^2\right)^n}\right)_{\rm DIV}(\vp)=
\sum_{j=0}^{n-3}{}'\,(-)^{j/2} A_n^j\left(\dl^{ab}\,(\nabla^2)^{j/2} +j\,\pa^a\pa^b\,(\nabla^2) ^{j/2-1}\right)\vp(0),
\ee
where the coefficients $A_n^j$ are those in \eref{boh} (for $j<n-2$), $\nabla^2=\pa_a\pa_a$, and the ``prime'' indicates that the sum is only over even $j$.

{\bf 2)  $n$ even.} If $n$ is even the procedure is exactly the same, but now in \eref{b4} we must keep also the term with $j=n-2$, that gives rise to a logarithmic divergence. For this term, before performing the rescaling $\vec r\ra \ve\vec r$ one must divide the integration region over $\vec r $ into, say, $0<r<1$ and $r>1$. The integral over $r>1$ converges as $\ve\ra 0$, while the one over $0<r<1$, after rescaling goes over to the region $0<r<1/\ve$. As $\ve\ra 0$ a logarithmic divergence arises then, coming  from large values of $r$. Instead of proceeding in \eref{b4} in this way, it can be seen that the correct divergence can also be obtained by considering the analytic continuation of $A_n^j$ with $j<n-2$ in \eref{boh}, and performing its expansion around $j=n-2$. One needs thus the expansion,
$$
{\G\left({n-j\over 2}-1\right)\over \ve^{n-j-2}}={2\over n-j-2}+\gamma-2\ln\ve+o(n-j -2)\quad\ra\quad-2\ln\ve +constant,
$$
where $\gamma$ is Euler's constant. With this replacement
from \eref{boh} we  obtain then,
\be\label{b6}
\left.A_n^j\right|_{j\ra n-2}\quad\ra\quad  {(-)^{n/2}\,\pi^{n/2}\over 2^{n-2}\,\G(n)\G\left({n\over2}\right)  }\cdot\ln\ve +\mbox{constant},
\ee
that reproduces $A_n^{n-2}$ in \eref{boh}.

We can thus take as general result the expression \eref{b5}, but with the sum extended up to $j=n-2$, and with the understanding that $A_n^{n-2}$ is given by the r.h.s. of \eref{b6}. Turning to abstract notation \eref{b5} translates then into,
\be\label{b6a}
\left({r^ar^b \over \left(r^2 +\ve^2\right)^n}\right)_{\rm DIV}=
\sum_{j=0}^{n-2}{}'\,(-)^{j/2}A_n^j\left(\dl^{ab}\,(\nabla^2)^{j/2} +j\,\pa^a\pa^b\,(\nabla^2) ^{j/2-1}\right)\dl^n(\vec r).
\ee
In generic flat coordinates, since on $\dl^n(\vec r)$ we have $-\nabla^2\dl^n= \Box \,\dl^n$, \eref{b6a} reads,
\be\label{b7}
\left({Q^{\m\a}\,Q^{\n\bt}\,x_\a x_\bt \over \left(Q_{\a\bt}\,x^\a x^\bt +\ve^2\right)^n}\right)_{\rm DIV}=
\sum_{j=0}^{n-2}{}'\,A_n^j\,\left(Q^{\m\n}\,\Box^{j/2}-j\,\pa^\m\pa^\n\,\Box ^{j/2-1}\right)\dl^n.
\ee
Applying \eref{b7} to the various terms of \eref{theg} it is straightforward to obtain \eref{divg}.

\subsection{Evaluation of a limit}\label{tlo}

In this section we determine the limit,
\be\label{b00}
{\rm Lim}_{\,\ve\ra 0}\,{\cal F}_\ve,
\ee
where the distributions ${\cal F}_\ve$ are given in \eref{feg},
\be\label{b8}
{\cal F}_\ve= -{\ve^2\over 2(Q_{\a\bt}x^\a x^\bt+\ve^2)^n}
+{1\over2}\,\sum_{j=0}^{n-2}{}'\,(n-j-2)\,A_n^j\,\Box^{j/2}\,\dl^n.
\ee
Notice that in the sum the term with $j=n-2$ drops out.
The determination of the distributional limit \eref{b00} requires to compute the ordinary limits $\lim_{\ve\ra0}{\cal F}_\ve(\vp)$.
Since in the second term in \eref{b8} the $\ve$--dependence is already explicit, see \eref{boh}, it is sufficient to concentrate on the first term. Proceeding as for
\eref{b3} and using static coordinates we obtain,
\ba
&&\left({\ve^2 \over \left(Q_{\a\bt}x^\a x^\bt +\ve^2\right)^n}\right)(\vp)=\ve^2
\int {\vp(\vec r)\,d^nr\over \left(r^2 +\ve^2\right)^n}\nn\\
&&=
{1\over \ve^{n-2}}\,\sum_{j=0}^{n-2}{}'\,{\ve^j\over j!} \left(\int {1\over (r^2+1)^n}\,
r^{a_1}\cdots r^{a_j}\,d^nr\right) \pa_{a_1}\cdots\pa_{a_j}\vp(0)+o(\ve). \nn
\ea
Using \eref{in1}, \eref{in2} to evaluate the integrals, one gets,
\be\label{b9}
\left({\ve^2 \over \left(Q_{\a\bt}x^\a x^\bt +\ve^2\right)^n}\right)(\vp)=
\sum_{j=0}^{n-2}{}'\,B_n^j\,\Box^{j/2}\,\vp(0) +o(\ve),
\ee
where,
\be\label{b10}
B_n^j= {(-)^{j/2}\,\pi^{n/2}\,\G\left({n-j\over 2}\right)\over 2^j\,\G(n)\,\G\left({j\over 2}+1\right)}\cdot{1\over \ve^{n-j-2}}.
\ee
With \eref{b9} from \eref{b8} we obtain,
\be\label{b11}
{\cal F}_\ve(\vp)= {1\over 2}\,\sum_{j=0}^{n-2}{}'\,\bigg[(n-j-2)\,A_n^j-B_n^j\bigg]\,\Box^{j/2}\,\vp(0) +o(\ve).
\ee
Comparing \eref{b10} with \eref{boh} we see that we have,
$$
(n-j-2)\,A_n^j=B_n^j, \quad \mbox{for}\quad0 \le j<n-2,
$$
while for $j=n-2$ this equality does not hold. Therefore, only the term with $j=n-2$ survives in the sum \eref{b11}. On the other hand, this term is there only for even $n$.  \eref{b11} gives thus,
\be\label{b12}
{\rm Lim}_{\,\ve\ra 0}\,{\cal F}_\ve=-{1\over 2}\,B_n^{n-2}\,\Box^{n/2-1}\,\dl^n=
{(-)^{n/2}\,\pi^{n/2}\over 2^{n-1}\,\G(n)\,\G\left({n\over 2}\right)}\,\Box^{n/2-1}\,\dl^n,\quad \mbox{for}\,\,n\,\,\mbox{even},
\ee
while for $n$ odd ${\rm Lim}_{\,\ve\ra 0}\,{\cal F}_\ve=0$. Using these results in \eref{limfeg} one obtains \eref{dtilde}.

\end{document}